\documentclass{jfm}

\usepackage{graphicx}
\usepackage{natbib}
\usepackage{subfigure}
\usepackage{rotating}
\ifCUPmtlplainloaded \else
  \checkfont{eurm10}
  \iffontfound
    \IfFileExists{upmath.sty}
      {\typeout{^^JFound AMS Euler Roman fonts on the system,
                   using the 'upmath' package.^^J}%
       \usepackage{upmath}}
      {\typeout{^^JFound AMS Euler Roman fonts on the system, but you
                   dont seem to have the}%
       \typeout{'upmath' package installed. JFM.cls can take advantage
                 of these fonts,^^Jif you use 'upmath' package.^^J}%
      }
  \else
  \fi
\fi

\ifCUPmtlplainloaded \else
  \checkfont{msam10}
  \iffontfound
    \IfFileExists{amssymb.sty}
      {\typeout{^^JFound AMS Symbol fonts on the system, using the
                'amssymb' package.^^J}%
       \usepackage{amssymb}%m,/m,/
       \let\le=\leqslant  \let\leq=\leqslant
       \let\ge=\geqslant  \let\geq=\geqslant
      }{}
  \fi
\fi

\ifCUPmtlplainloaded \else
  \IfFileExists{amsbsy.sty}
    {\typeout{^^JFound the 'amsbsy' package on the system, using it.^^J}%
     \usepackage{amsbsy}}
    {}
\fi

\newsavebox{\astrutbox}
\sbox{\astrutbox}{\rule[-5pt]{0pt}{20pt}}

\title[Viscous boundary layer in turbulent thermal convection: the effect of cell tilting]{Viscous boundary layer properties in turbulent thermal convection in a cylindrical cell: the effect of cell tilting}

\author[Ping Wei and Ke-Qing Xia]%
{P\ls I\ls N\ls G\ns W\ls E\ls I\ns  A\ls N\ls D\ns K\ls E\ls -\ls Q\ls I\ls N\ls G\ns X\ls I\ls A\ns}

\affiliation{Department of Physics, The Chinese University of Hong Kong, Shatin, China}

\pubyear{}
\volume{}
\pagerange{}
\date{?; revised ?; accepted ?. - To be entered by editorial office}
\begin{document}

\maketitle

\begin{abstract}
We report an experimental study of the properties of the velocity boundary layer in turbulent Rayleigh-B\'{e}nard convection in a cylindrical cell. The measurements were made at Rayleigh numbers $Ra$ in the range $2.8\times10^{8}<Ra<5.6\times10^{9}$ and were conducted with the convection cell tilted with an angle $\theta$ relative to gravity, at $\theta=0.5^{o}$, $1.0^{o}$, $2.0^{o}$, and $3.4^{o}$, respectively. The fluid was water with Prandtl number $Pr=5.3$.

It is found that at small tilt angles ($\theta \le 1^{o}$), the measured viscous boundary layer thickness $\delta_{v}$ scales with the Reynolds number $Re$ with an exponent close to that for a Prandtl-Blasius laminar boundary layer, i.e. $\delta_{v} \sim Re^{-0.46\pm0.03}$. For larger tilt angles, the scaling exponent of $\delta_{v}$ with $Re$ decreases with $\theta$. The normalized mean horizontal velocity profiles measured at the same tilt angle but with different $Ra$ are found to have an invariant shape. But for different tilt angles, the shape of the normalized profiles is different.

It is also found that the Reynolds number $Re$ based on the maximum mean horizontal velocity scales with $Ra$ as $Re \sim Ra^{0.43}$ and the Reynolds number $Re_{\sigma}$ based on the maximum rms velocity scales with $Ra$ as $Re_{\sigma} \sim Ra^{0.55}$, with both exponents do not seem to depend on the tilt angle $\theta$.

Several wall quantities are also measured directly and their dependency on $Re$ are found to agree well with those predicted for a classical laminar boundary layer. These are the wall shear stress $\tau$ ($\sim Re^{1.46}$), the viscous sublayer $\delta_{w}$ ($\sim Re^{0.75}$), the friction velocity $u_{\tau}$ ($\sim Re^{-0.86}$) and the skin friction coefficient $c_{f}$ ($\sim Re^{-0.46}$). Again, all these near-wall quantities do not seem to depend on the tilt angle. 

We also examined the dynamical scaling method proposed bys Zhou and Xia [{\it Phys. Rev. Lett.} {\bf 104}, 104301 (2010)] and found that in both the laboratory and the dynamical frames the mean velocity profiles show deviations from the theoretical Prandtl-Blasius profile, with the deviations increase with $Ra$. But profiles obtained from dynamical scaling in general have better agreement with the theoretical profile. It is also found that the effectiveness of this method appears to be independent of $Ra$.

\end{abstract}
%\tableofcontents

\section{Introduction}\label{sec:introduction}
\subsection{Rayleigh-B\'{e}nard convection}\label{subsec:convection}

Rayleigh-B\'{e}nard (RB) convection, which is a fluid layer heated from below and cooled from the top, is an idealized model to study turbulent flows involving heat transport and has attracted much attention during the past few decades \citep*{siggia1994arfm,kadanoff2001pt,agl2009rmp,lx2010arfm}. 
The system is characterized by two control parameters: the Rayleigh number $Ra$, Prandtl number $Pr$, which are defined as  
\begin{equation}
Ra=\frac{\alpha g \Delta T H^{3}}{\nu\kappa}, 
\end{equation}
and
\begin{equation}
Pr=\frac{\nu}{\kappa},
\end{equation}
respectively. Here $\alpha$ is the thermal expansion coefficient, $g$ the gravitational acceleration, $\Delta T$ the temperature difference between the bottom and the top plates, $H$ the height of the fluid layer between the plates, $\nu$ the kinematic viscosity, and $\kappa$ the thermal diffusivity of the convecting fluid. In addition, the aspect ratio $\Gamma=D/H$ ($D$ is the lateral dimension of the system) also plays an important role in the structures and dynamics of the flow.
 
In a fully developed Rayleigh-B\'{e}nard turbulent flow,
most of the imposed temperature difference is localized in the thermal boundary layers near the surface of the top and bottom plates, within which heat is transported via conduction \citep*{wu1991pra,belmonte1994pre,liu1998pre}. The velocity field has the same character: velocity gradient is localized in a thin layer near the plates, which is called viscous boundary layer. Turbulent flow in the central region of the RB cell is approximately homogenous and isotropic\citep*{zhou2008jfm,ni2011prl,ni2012jfm}. As the top and bottom boundary layers contribute the main resistance to heat transfer through the cell and thus dominantly determine the Nusselt number, they deserve special attention. Indeed, nearly all theories in RB convection are in essence boundary-layer (BL) theories. For example, a turbulent BL was assumed in the early marginal stability theory \citep*{malkus1954prsl} and also in the models by \citet*{shraiman1990pra} and \citet{siggia1994arfm} and by \citet*{dubrulle2001epjb,dubrulle2002epjb}. On the other hand, a Prandtl-Blasius (PB)  type laminar BL was assumption in the  Grossmann \& Lohse (GL) theory \citep*{gl2000jfm,gl2001prl,gl2002pre,gl2004pof}. Therefore, direct characterization of the BL properties is essential for testing and differentiating the various theoretical models, and will also provide insight into the physical nature of turbulent heat transfer.

\subsection{Boundary layer measurements in turbulent thermal convection}\label{subsec:boundary layer measurement}

One of the earlier measurements of temperature and also velocity profiles in turbulent RB convection was taken by \citet*{tilgner1993pre} in water ($Pr=6.6$) at the fixed $Ra=1.1\times10^{9}$ and at a fixed lateral position. \citet*{belmonte1993prl} extended these measurements over the range $5\times10^{5}\leq Ra\leq 10^{11}$ in compressed gas (air) at room temperature (Pr=0.7), but still at fixed lateral position. \citet{liu1998pre} measured the mean temperature profiles at various horizontal positions on the lower plate of a cylindrical convection cell, the result shows that the thermal layer thickness $\delta_{th}$ varied over the plate for the same $Ra$, and the thinnest BL is closed to the center of the plate. \citet*{wang2003epjb} found similar results for a cubic cell. 
\citet{dupuits2007jfm} measured high-resolution temperature profiles in Rayleigh-B\'{e}nard convection near the top plate of a cylindrical container with air ($Pr=0.7$) as the working fluid. Their result shows that the thermal BL thickness $\delta_{th}\sim Ra^{-0.25}$ in the cell with $\Gamma=1.13$. \citet*{sun2008jfm} found that the thermal boundary layer thickness  scales with $Ra^{-0.32}$ in a rectangular cell, at $Pr=4.3$ and $Ra$ ranging from $10^{8}$ to $10^{10}$.

For the velocity measurement, the methods for determining the velocity profiles near the solid walls of the cell are developed in recent years. Since strong temperature fluctuations exist in Rayleigh-B\'{e}nard convection, the well-established hot-wire anemometry could not be applied to this system. For the viscous boundary layer, the large temperature fluctuations make conventional laser Doppler velocimetry ineffective because the temperature fluctuations cause fluctuations in the refractive index of the fluid that in turn make it difficult to steadily focus two laser beams to cross each other in the fluid \citep*{xia1995josa}. \citet{tilgner1993pre} introduced an electrochemical labeling method and measured the velocity profile and boundary layer thickness near the top plate of a cubic cell filled with water, but only at a single value of $Ra$. In a later study, \citet*{belmonte1993prl,belmonte1994pre} developed an indirect method --- the correspondence between the peak position of the cutoff frequency profile of the temperature power spectrum and the peak position of the velocity --- to infer the viscous boundary boundary layer thickness in gaseous convection. This method has subsequently been used to infer the viscous layer in thermal convection in mercury \citep*{sano1997pre}. It may just be that the method works in certain situations, but there is no theoretical basis for it. \citet*{xin1996prl}, using a novel light-scattering technique developed by \citet{xia1995josa}, conducted the first direct systematic measurement of velocity profiles in RB convection  in a cylindrical cell as a function of $Ra$. They found $\delta_{\upsilon}\sim Ra^{-0.16}$ from the velocity profile above the center of the lower plate. \citet{qiu1998pre_b,qiu1998pre} extended these measurements to convection in cubic cells, finding the same scaling exponent $-0.16$ at the bottom plate, but at the sidewall a different result $\delta_{\upsilon}\sim Ra^{-0.26}$. Using various organic liquids, \citet{lam2002pre} explored the $Pr$ dependence, finding $\delta_{\upsilon}\sim Pr^{0.24}Ra^{-0.16}$. With the measured $Ra-Re$ scaling relationship $Re \sim Ra^{0.5}$ obtained in these studies (in fact, it was the Peclet number $Pe = \upsilon L/\kappa$ rather than $Re$ in some of these studies; please see \citet*{sun2005pre_b} for more detailed discussions), the above results imply a scaling relation $\delta_{\upsilon}\sim Re^{-0.32}$. In recent years, the technique of particle image velocimetry (PIV) has been introduced to the experimental study of thermal convection \citep*{xia2003pre,sun2005prl, sun2005pre,sun2005pre_b}.  \citet{sun2008jfm} further applied the PIV technique to study the viscous BL in a rectangular cell. Their results show that $\delta_{\upsilon}\sim Ra^{-0.27}$ and $\delta_{\upsilon} \sim Re^{-0.5}$, which showed that the viscous BL in thermal turbulence has the same $Re$-scaling as a Prandtl-Blasius laminar BL. This result validates the laminar BL assumption made in the GL model in a scaling sense. Thus it appears there is a discrepancy in the measured scaling exponent of $\delta_{\upsilon}$ with respect to $Ra$ ($Re$) between those obtained in cylindrical and cubic cells and that obtained in rectangular cells. In \citet{sun2008jfm} it was argued that because of the more complicated flow dynamics of the large-scale circulation (LSC), such as the azimuthal motion in the cylindrical cell \citep{sun2005pre,brown2005prl,Xi2006pre} and the secondary flows in the cubic cell \citep{qiu1998pre_b}, the shear flow near the plates is less steady as compared to that in the rectangular cell which is more close to quasi-two-dimension (quasi-2D). As the viscous boundary is  created by the shear of the LSC, this may plausibly change the BL properties, resulting in a different exponent.  However, the above argument has not been substantiated experimentally. Part of the motivation of the present work is to determine how the three-dimensional LSC dynamics will affect the BL properties.   It is known that titling the cell by a small angle will ``lock'' the LSC in a fixed azimuthal plane in the sense that it will limit the range of the LSC's azimuthal meandering \citep{sun2005prl} and reduce its azimuthal oscillation amplitude near the top and bottom plates of the cell \citep*{ahlers2006jfm}.  In the present work, we present measurements of BL properties in a cylindrical cell with the cell titled over a range of angles. For small titling angles, we measure a boundary layer under a more steady shear comparing to the ``leveled"   case when the LSC can freely meander in the azimuthal direction but presumably the BL is otherwise unperturbed under such a small titling angle ($\le 1^{o}$). We  also examine how the BL scaling exponent and other BL properties behave when the titling angle becomes not so small ($\ge 1^{o}$), which would amount to a perturbation to the BLs. Boundary layers play such an important role in turbulent thermal convection, it is therefore important to examine how BLs respond to external perturbations. Understanding the stability or instability of BLs is also relevant to the search for the so-called ultimate state of thermal convection, as the transition from the ``classical state" to the ultimate one is essentially an instability transition of the BL from being laminar to being turbulent.

In addition to the scaling of the BL thickness, the shape of the velocity profile near the top and bottom plates has attracted a lot of attention recently. Although the BL has been found scaling wise to be of Prandtl-Blasius type  (at least in the quasi-2D case), the time-averaged velocity profiles are found to differ with the theoretically predicted one \citep{dupuits2007prl,sun2008jfm}, especially for the region around the thermal BL. Recently, \citet*{zhou2010prl} have proposed a dynamic scaling method that shows that the mean velocity profile measured in the laboratory frame can be brought into coincidence with the theoretical Prandtl-Blasius laminar BL profile, if it is resampled relative to the time-dependent frame that fluctuates with the instantaneous BL thickness. This method was tested initially for the case of velocity profile in turbulent convection in a quasi-2D rectangular cell with water as working fluid ($Pr = 4.3$). In a follow-up study using two-dimensional DNS data,  \citet{zhou2010jfm} found that the method is also valid for thermal boundary layers and for the case of $Pr =0.7$ as well. More recently, these authors further shown, again using numerical data, that the method works also in other positions in the horizontal plate other than the central axis \citep{zhou2011pof} and in three-dimension (3D) cylindrical cell for moderate values of $Ra$ \citep{stevens2012pre}. However, \citet*{scheel2012jfm} and \citet*{schumacher2012jfm}, both using numerical approaches, have found that dynamic scaling works less well in the 3D cylindrical geometry than in the quasi-2D case. However, the method has not been tested experimentally so far in a 3D system. Here we would like to  examine the dynamical scaling method using the experimentally obtained instantaneous velocity profiles in our three-dimensional cylindrical cell.

\subsection{Organization of the paper}\label{subsec:Organization}

The remainder of this paper is organized as follows. We give detailed descriptions of the experimental setup and measurement instrumentation in \S\ref{sec:apparatus} and present and analyze experimental results in \S\ref{sec:result}, which are divided into six subsections. In \S\ref{subsec:ThermalBL} we present the measured temperature profiles and corresponding position-dependent fluid properties, which will be used to calculate the viscous and Reynolds stresses. In \S\ref{subsec:Mean velocity profiles}, the measured velocity profiles and their characterizations are presented. In \S\ref{subsec:Scaling for Ra-scan}, the scaling properties, with both $Ra$ and $Re$, of the thickness $\delta_{v}$ obtained from the mean velocity profiles and $\delta_{\sigma}$ obtained form r.m.s. velocity profiles, are presented and discussed. We also discuss the influence of the cell tilting angle $\theta$ on the boundary layer scaling. In \S\ref{subsec:statistical properties of the velocity field in the boundary layer} statistical properties (r.m.s. and skewness) of the velocity field in the boundary layer region are discussed. In \S\ref{subsec:the quantities of turbulence} we present results of the viscous and Reynolds shear stresses distributions in the boundary layer, and discuss the scaling of the wall quantities. In \S\ref{subsec:Instantaneous boundary layer properties} we test the dynamic scaling method with respect to the measured instantaneous velocity profiles. We summarize our findings and conclude in \S\ref{sec:summary}.

\section{Experimental apparatus}\label{sec:apparatus}
\subsection{Convection cell} \label{subsec:Cell}

\begin{figure}
\centerline{
\includegraphics[width=\textwidth]{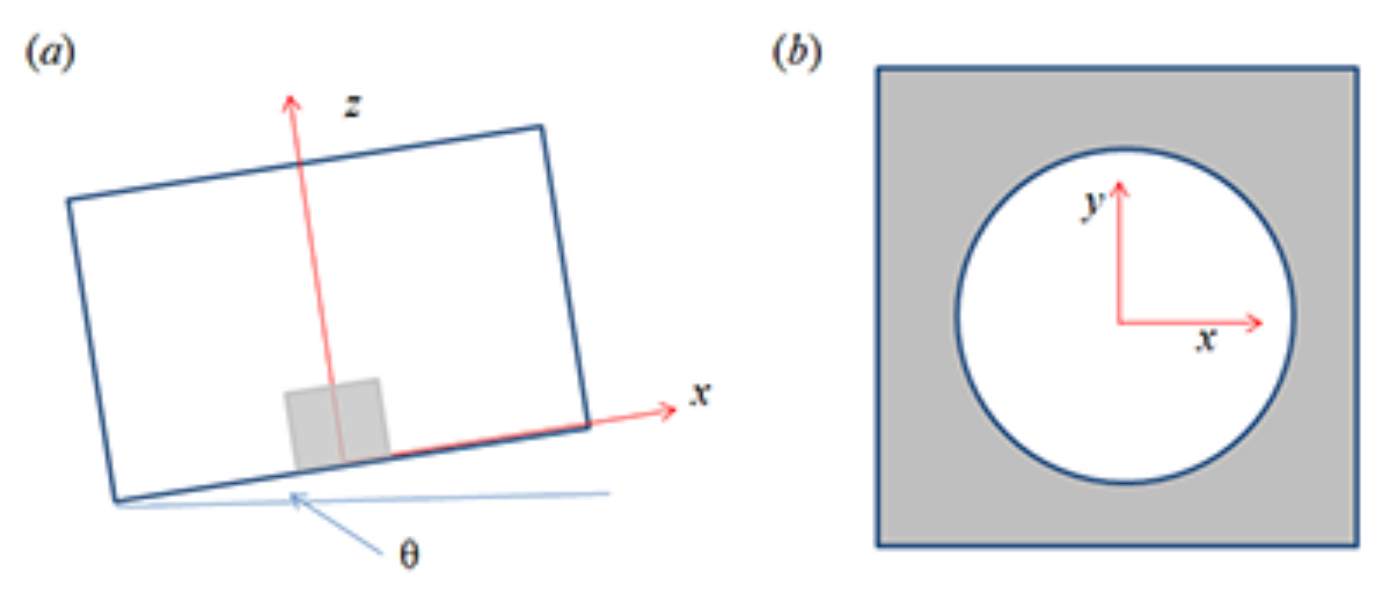}
}
\caption{\label{fig:SetupForPIV} Sketch of the convection cell and the Cartesian coordinates used in temperature and velocity measurements. (a) side view of the setup, and (b) the top view.}
\end{figure}

The measurements were made in a cylindrical Rayleigh-B\'{e}nard  convection cell, which has been described in detail previously \citep{zhou2002prl,sun2005pre,ni2011pof}. 
Here we give only its essential features. The top and bottom conducting plates are made of pure copper with a thin layer of nickel to avoid oxidation. The sidewall is made of Plexiglas. To avoid distortions in the images viewed by the camera, a square-shaped jacket is fitted around the sidewall of the convection cell. As shown in figure~\ref{fig:SetupForPIV}(b), the jacket is filled with water. The diameter and height of the cell is $D=19.6$ cm and $H=18.6$ cm, respectively. The aspect ratio $\Gamma = D/H$ is thus close  to $1$. Two (three) thermistors are embedded in the top (bottom) plate. The top plate temperature is maintained constant by a refrigerated circulator (Polyscience Model 9702) that has a temperature stability of $0.01^{o}C$. A NiChrome wire (26 Gauge, Aerocon Systems) surrounded by fiberglass sleeving and Teflon tape is distributed inside the grooves carved under the bottom plate. The wire is connected with five DC power supplies (GE Model GPS-3030) in series to provide constant and uniform heating. During the measurement, the whole cell is placed in a homemade thermostat box that is kept at the same temperature ($30^{o}C$) as that of the fluid at the centre of the cell. During the experiment the cell was tilted by an angle $\theta$ such that the circulation plane of the LSC was parallel to the image plane of the camera (the $x$-$z$ plane, see figure 1). 

\subsection{PIV measurement} \label{subsec:PIV}

The application of PIV to thermal turbulence has been described in detail in several previous publications \citep{xia2003pre,sun2005pre,sun2008jfm}. Here we only provide details concerning the particular features of the present experiment. The PIV system consists of one CCD camera with  $2048 \times 2048$ pixels, a dual pulse Nd-YAG laser with $135$ mJ per pulse, a synchronizer and software.
As the cell was titled, both the CCD and the laser light-sheet were titled accordingly with the same angle. A 105 mm focal-length macro lens was attached to the CCD to achieve a measuring area with size varying from $18\times18$ mm$^{2}$ to $30\times30$ mm$^{2}$. 
Each 2D velocity vector is calculated from a subwindow (32 pixels $\times$
32 pixels) that has 50$\%$ overlap with its neighboring subwindows, so each vector corresponds to a region of 16 pixels $\times$ 16 pixels and each velocity map contains $127 \times 127$ velocity
vectors in the $x$-$z$ plane (see figure 1). 
 This corresponds to  spatial resolutions of about $0.135\times0.135$ mm$^{2}$ to $0.236\times 0.236$ mm$^{2}$  for velocities $u$ and $w$ measured in the horizontal $x$ and vertical $z$ directions, respectively.   For the measurement at $\theta=3.4^{o}$  particles with diameter $2~\mu$m were used, while particles with  diameter of $10~\mu$m were used for measurements with other three tilted angles. For each run, typically about 25200 image pairs were acquired with frame rate of $2$ Hz.

\begin{figure}
\centerline{
\includegraphics[width=\textwidth]{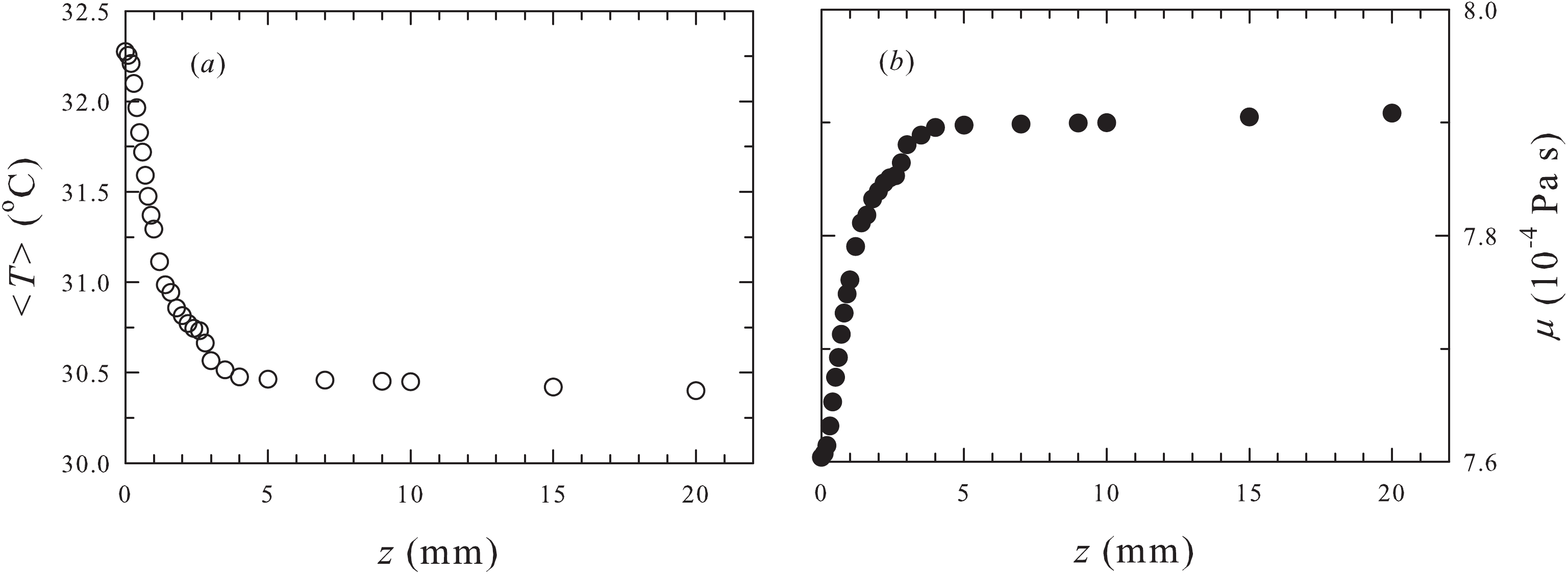}
}
\caption{(a) A profile of mean temperature $\langle T\rangle$ measured at $Ra=6.8\times10^{8}$ and $\theta=3.4^{o}$. (b) A profile of dynamic viscosity $\mu$ obtained from the mean temperature profile in (a).}
\label{fig:ThermalBL}
\end{figure}

\section{Results and discussion} \label{sec:result}

PIV measurements were made at four values of the titling angle $\theta$ $=0.5^{o}$, $1^{o}$, $2^{o}$, and $3.4^{o}$. For each $\theta$, measurements over a range of $Ra$ were made. Table~\ref{table:parameters} lists the parameters ($\theta$, $Ra$ and $Pr$) of each measurement, which typically lasted for about 3.5 hours. As already mentioned, titling the cell by a small angle has the effect of ``locking" the LSC's circulation plane at a fixed azimuthal angle (in reality it restricts the angular range of the LSC's azimuthal meandering). Thus, measurements made with small $\theta$ are aimed at studying BL properties under more steady shear, but the BL itself is  assumed to be unperturbed otherwise. For large values of $\theta$ we wish to examine how the BL responds to relatively large perturbations. 

\begin{table}
  \begin{center}
\def~{\hphantom{0}}
  \begin{tabular}{cccccccccc}
$ \theta$	&		Ra				&	Pr	&	$U_{max}$	&	$\delta_{v}$	&	$ \theta$	&		Ra				&	Pr	&	$U_{max}$	&	$\delta_{v}$			\\
$ (deg.)$	&						&		&	$ (mm/s)$	&	$ (mm)$	&	$ (deg.)$	&						&		&	$ (mm/s)$	&	 $(mm)$			\\
$0.5$	&	$	5.77	\times	10^{8}	$	&	5.45 	&	4.77 	&	2.80 	&	$2.0$	&	$	1.34	\times	10^{9}	$	&	5.42 	&	7.47 	&	2.57 			\\
	&	$	2.79	\times	10^{8}	$	&	5.41 	&	3.57 	&	3.40 	&		&	$	5.85	\times	10^{8}	$	&	5.42 	&	5.21 	&	3.25 			\\
	&	$	1.55	\times	10^{9}	$	&	5.39 	&	7.38 	&	2.44 	&		&	$	5.53	\times	10^{9}	$	&	5.32 	&	13.49 	&	1.97 			\\
	&	$	2.93	\times	10^{9}	$	&	5.39 	&	9.11 	&	2.13 	&	$3.4$	&	$	2.74	\times	10^{8}	$	&	5.67 	&	4.02 	&	4.54 			\\
	&	$	4.26	\times	10^{9}	$	&	5.38 	&	10.88 	&	2.06 	&		&	$	4.19	\times	10^{8}	$	&	5.66 	&	4.70 	&	4.23 			\\
$1.0$	&	$	1.68	\times	10^{9}	$	&	5.45 	&	7.53 	&	3.46 	&		&	$	6.69	\times	10^{8}	$	&	5.66 	&	6.00 	&	3.58 			\\
	&	$	3.19	\times	10^{9}	$	&	5.39 	&	9.96 	&	2.87 	&		&	$	1.38	\times	10^{9}	$	&	5.68 	&	7.59 	&	2.70 			\\
	&	$	5.54	\times	10^{9}	$	&	5.42 	&	11.60 	&	2.66 	&		&	$	2.19	\times	10^{9}	$	&	5.64 	&	9.40 	&	2.50 			\\
	&	$	9.46	\times	10^{8}	$	&	5.45 	&	5.72 	&	3.82 	&		&	$	2.66	\times	10^{9}	$	&	5.66 	&	10.20 	&	2.11 			\\
	&	$	2.40	\times	10^{8}	$	&	5.42 	&	3.06 	&	5.07 	&		&	$	9.89	\times	10^{8}	$	&	5.66 	&	6.47 	&	3.00 			\\
	&	$	6.00	\times	10^{8}	$	&	5.43 	&	4.71 	&	3.94 	&		&	$	3.20	\times	10^{9}	$	&	5.65 	&	10.84 	&	2.14 			\\
$2.0$	&	$	2.76	\times	10^{8}	$	&	5.41 	&	3.76 	&	4.12 	&		&	$	4.28	\times	10^{9}	$	&	5.63 	&	12.65 	&	1.89 			\\
	&	$	2.78	\times	10^{9}	$	&	5.42 	&	10.09 	&	2.15 	&		&	$	5.19	\times	10^{9}	$	&	5.56 	&	13.49 	&	1.84 			\\

  \end{tabular}
  \caption{Control parameters of the experiment: cell tilt angle $\theta$, the Rayleigh number $Ra$ and the Prandtl number $Pr$;  and the measured maximum horizontal velocity $U_{max}$ and viscous boundary layer thickness $\delta_{v}$. The data are listed in chronological order.}
  \label{table:parameters}
  \end{center}
\end{table}

\subsection{Temperature profile and fluid properties}\label{subsec:ThermalBL}

The local values of fluid properties are needed in calculating the viscous and Reynolds shear stresses, which requires measurement of the local temperature.  Temperature profiles for the leveled case have been measured systematically by \citet{liu1998pre} in a similar cylindrical cell. To check whether titling the cell by a relatively large angle will change the temperature profile, we measured one mean temperature along the central axis ($x=y=0$) of the cell at a titling angle $\theta=3.4^{o}$ ($Ra=6.8\times10^{8}$). The result is shown in figure~\ref{fig:ThermalBL}(a) and the dynamic viscosity corresponding to the local temperature is  shown in figure~\ref{fig:ThermalBL}(b). As these results are similar to those obtained in previous studies  by \citet{liu1998pre} and \citet{sun2008jfm}, we will use results from those studies at similar $Ra$ in the calculations of Reynolds stress (Sec.~\ref{subsec:the quantities of turbulence}) and other wall quantities that require position dependent viscosity (density).

\subsection{Velocity profiles and the Reynolds number scaling}\label{subsec:Mean velocity profiles}

\begin{figure}
\begin{minipage}[t]{0.5\linewidth}
\centering
\includegraphics[width=1\textwidth]{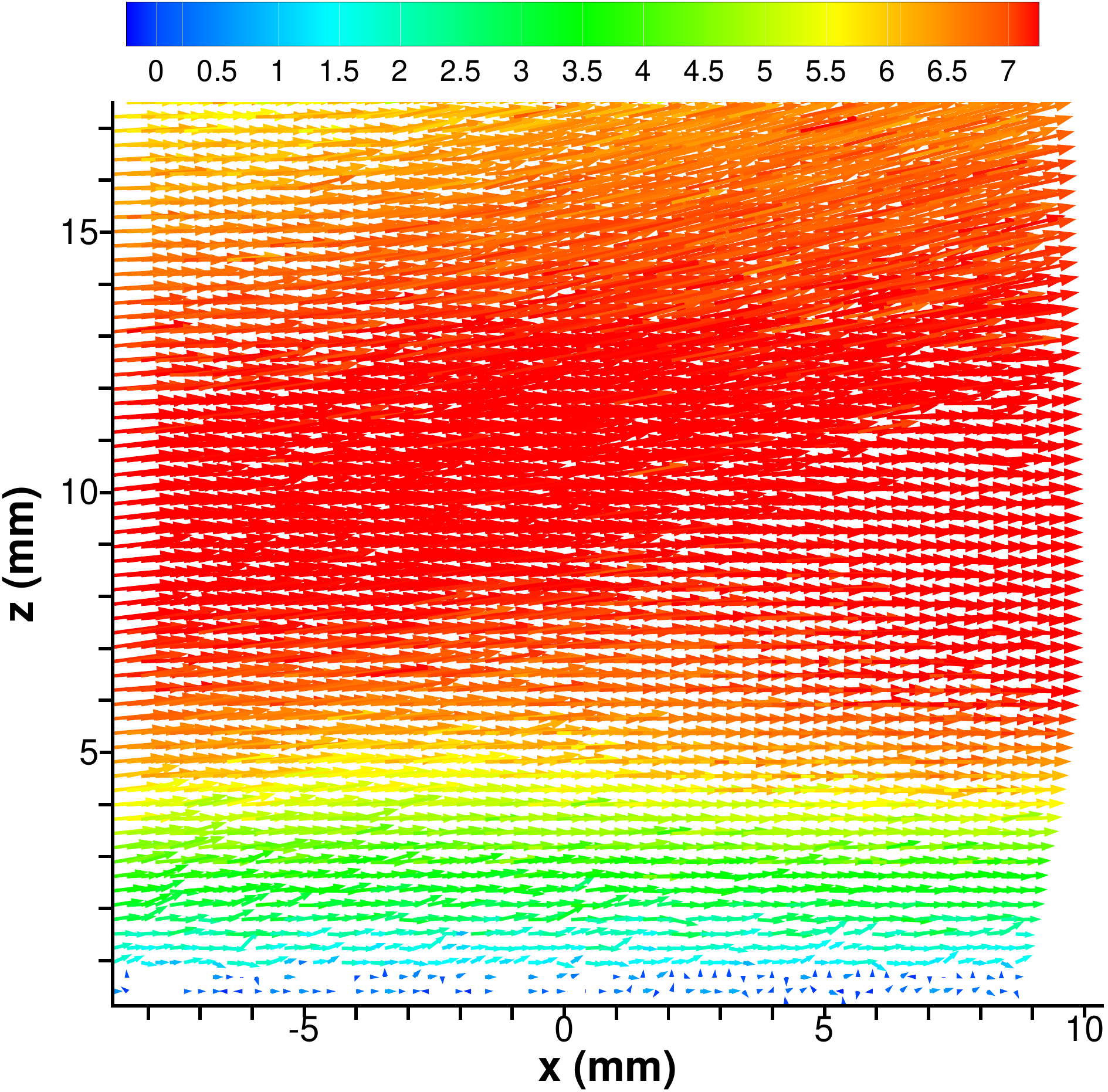}
\end{minipage}%
\begin{minipage}[t]{0.5\linewidth} 
\centering
\includegraphics[width=1\textwidth]{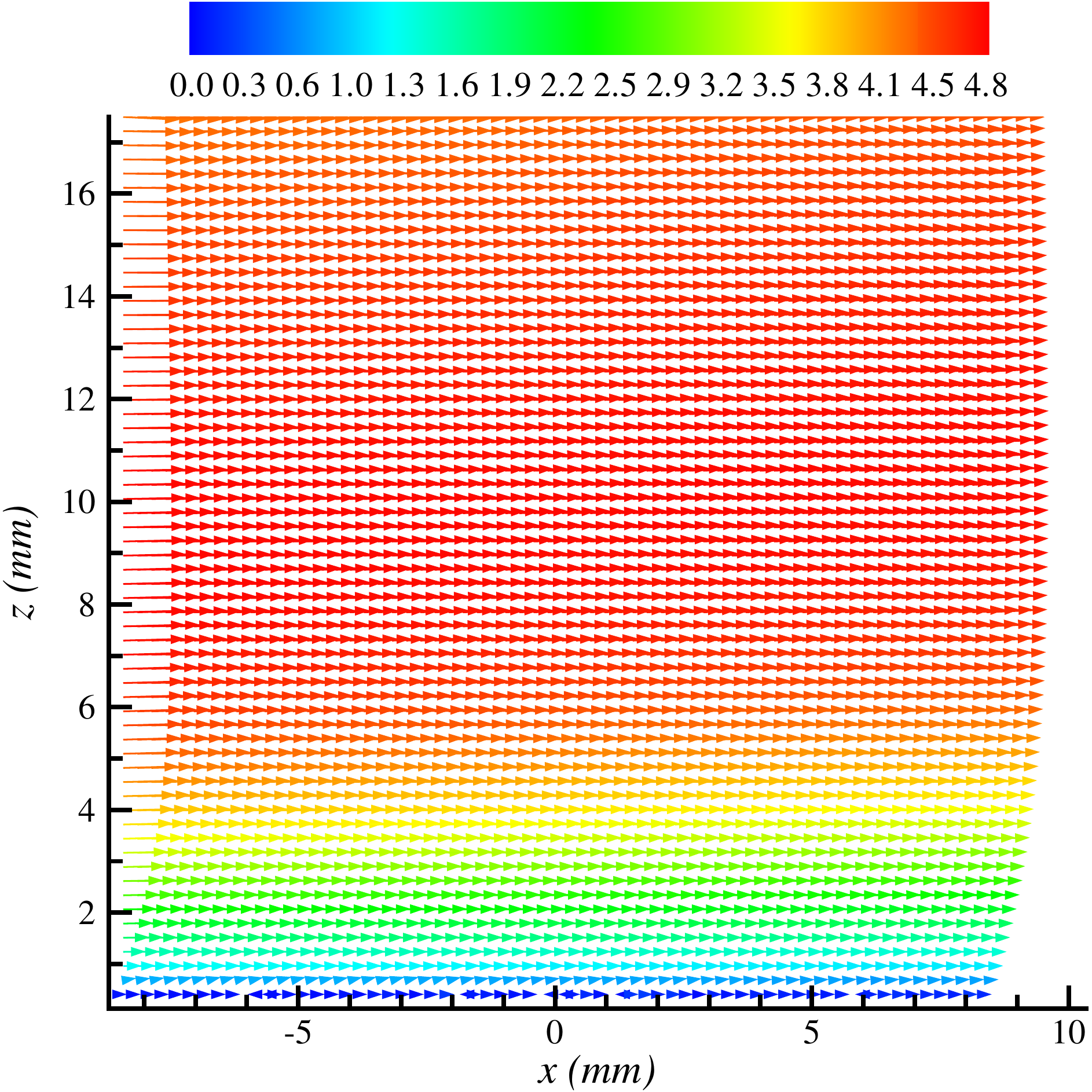}
\end{minipage}
\caption{ Coarse-grained vector maps of the instantaneous (a) and time-averaged (b) velocity field measured near the center of the bottom plate ($Ra=4.2\times10^{8}$ with $\theta=3.4^{o}$), the velocity scale bar is in unit of $mm/s$.}
\label{fig:AveragedField}
\end{figure}

\begin{figure}
\centerline{
\includegraphics[width=\textwidth]{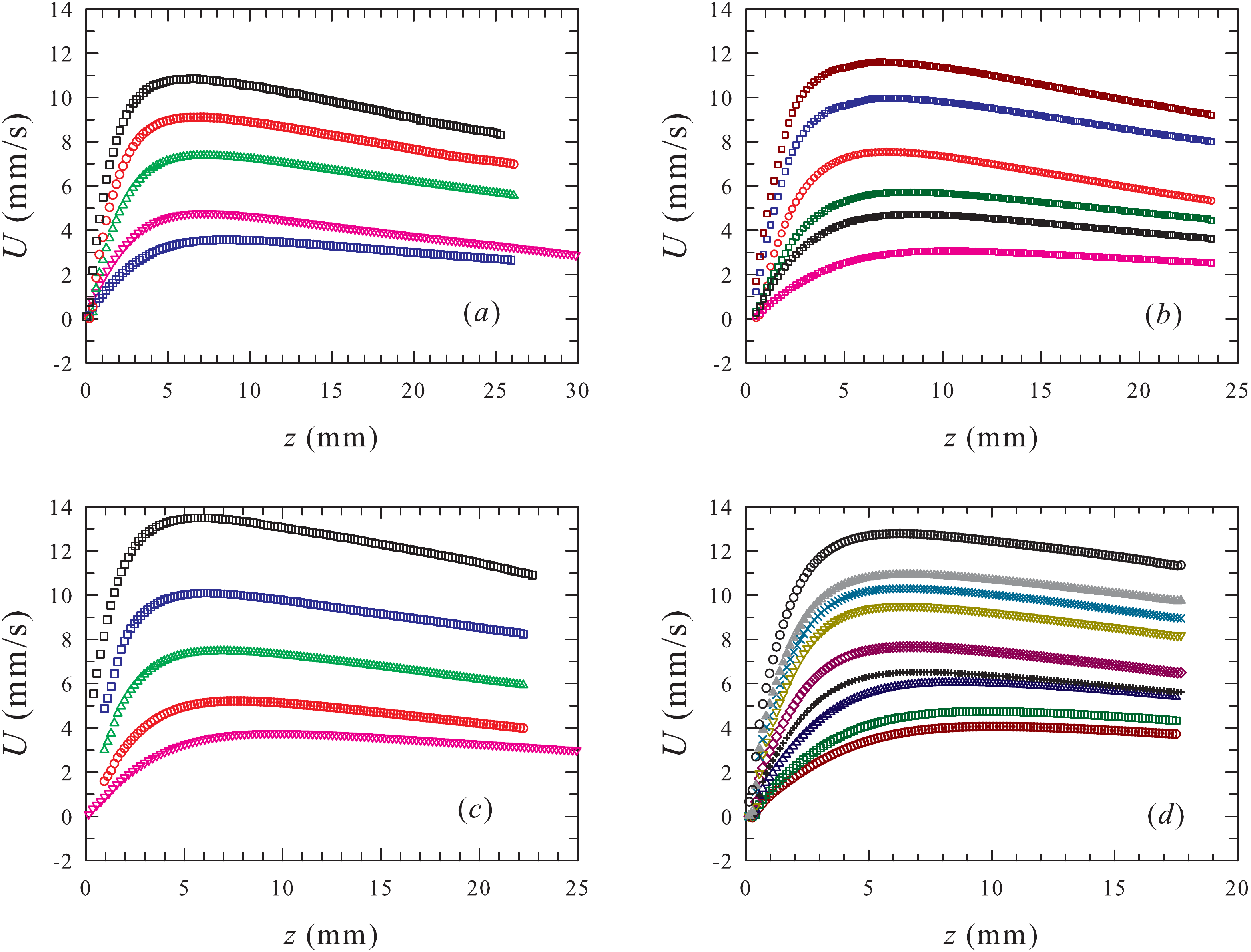}
}
\caption{\label{fig:UnNormalizedU} Time-averaged horizontal velocity profiles measured at tilt angles $\theta= 0.5^{o}$ (a), $1.0^{o}$ (b), $2.0^{o}$ (c)  and $3.4^{o}$ (d). In each plot the corresponding value of $Ra$ decreases from top to bottom (see Table~\ref{table:parameters} for exact values).}
\end{figure}

\begin{figure}
\centerline{
\includegraphics[width=\textwidth]{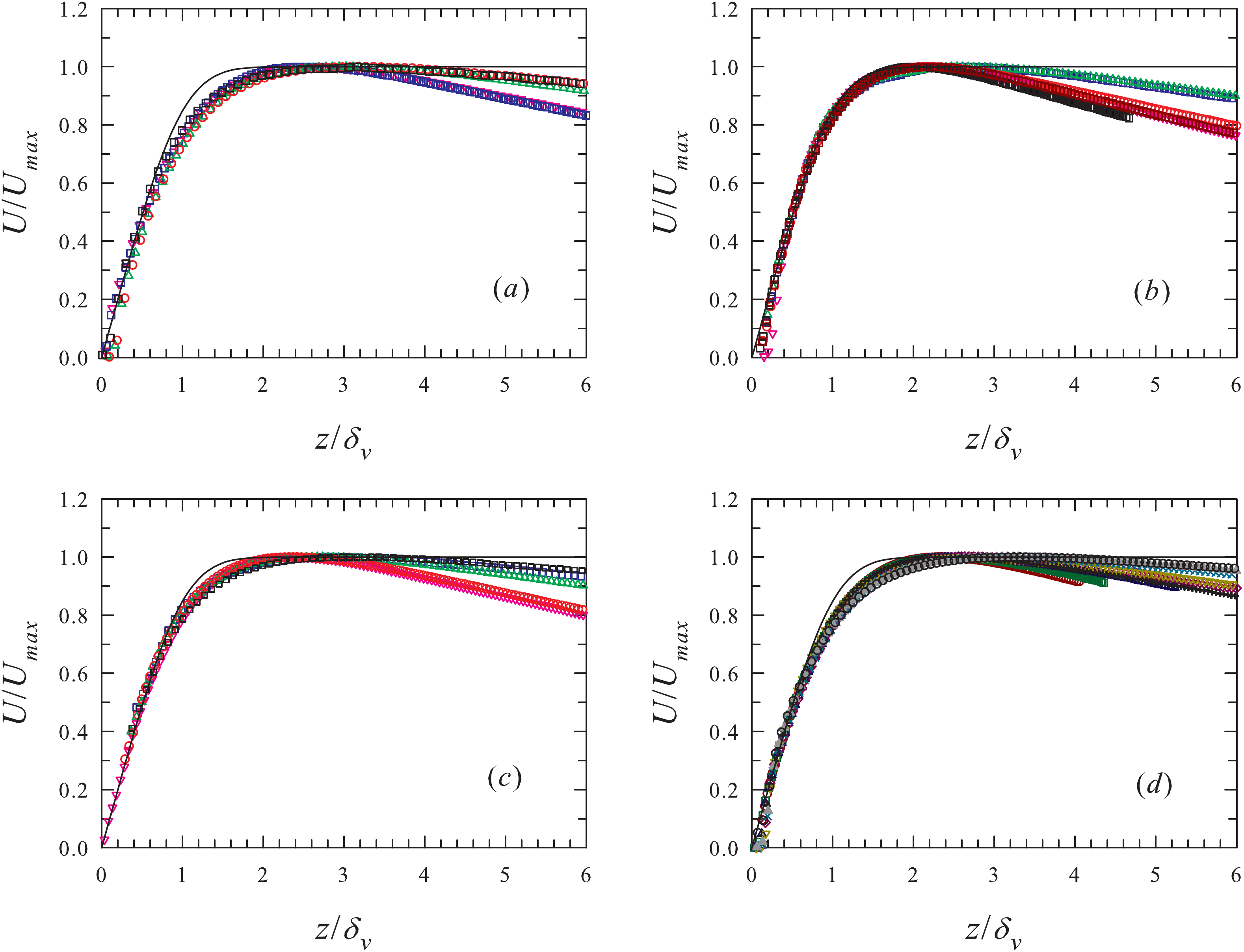}
}
\caption{\label{fig:NormalizedU} Profiles normalized by their respective maximum velocity $U_{max}(Ra)$ and the corresponding viscous boundary layer thickness $\delta_{v}(Ra)$ with tilt angles $\theta= 0.5^{o}$ (a), $1.0^{o}$ (b), $2.0^{o}$ (c)  and $3.4^{o}$ (d). The solid line in each plot represents the theoretical Prandtl-Blasius profile.}
\end{figure}

\begin{figure}
\centerline{
\includegraphics[width=0.7\textwidth]{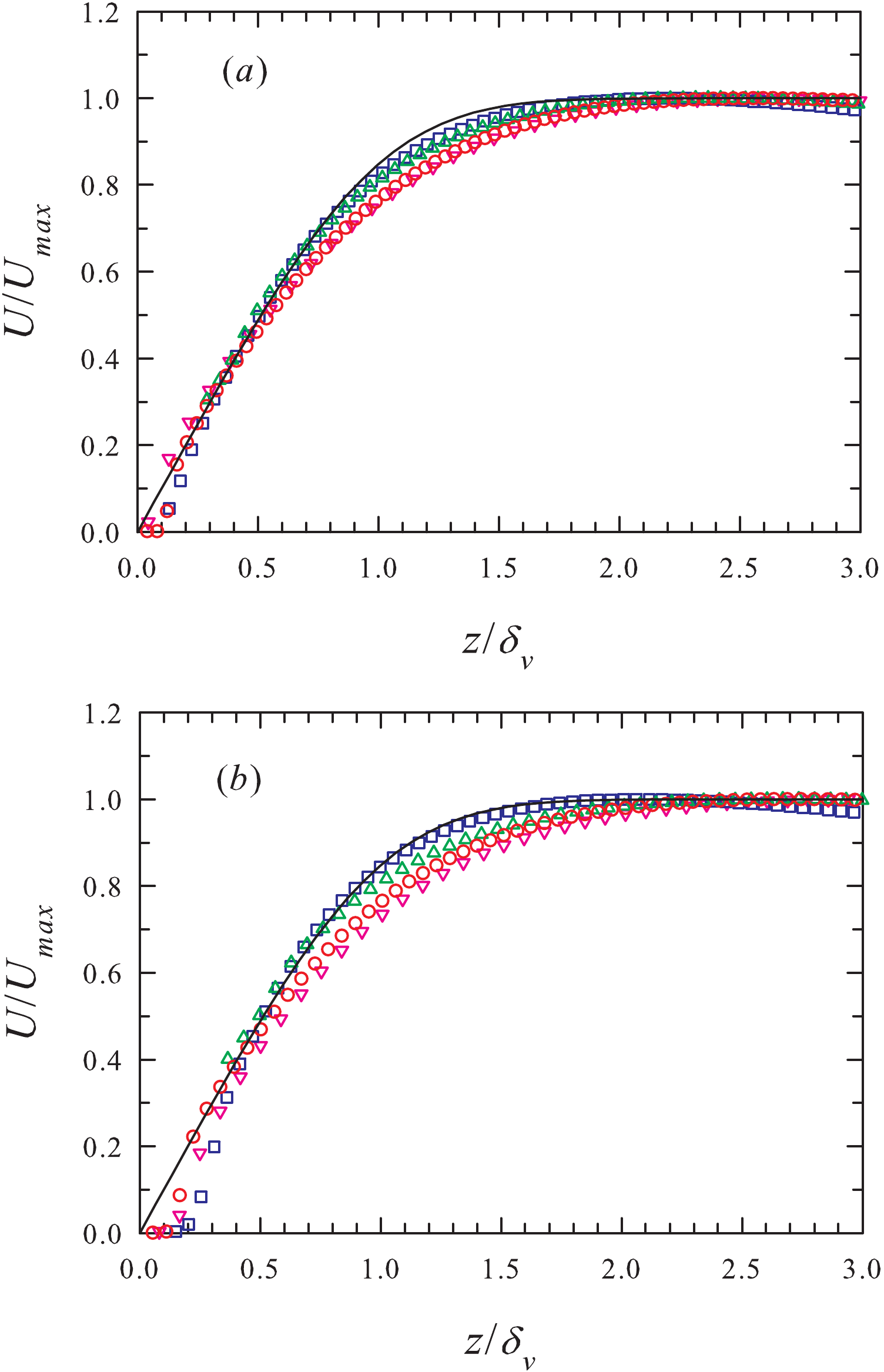}
}
\caption{\label{fig:Rescale-TwoRa} Normalized profiles measured at different tilt angles $\theta$ but with approximately the same value of $Ra$. Profiles in (a) have a nominal value of $Ra = 5\times 10^{8}$ and in (b) have a nominal value of $Ra = 1.5\times 10^{9}$. In both figures the symbols are: inverted triangles ($\theta = 0.5^{o}$); squares ($\theta = 1.0^{o}$); triangles ($\theta = 2.0^{o}$); and circles ($\theta = 3.4^{o}$).  The solid line in each plot represents the theoretical Prandtl-Blasius profile.}
\end{figure}

\begin{figure}
\centerline{
\includegraphics[width=\textwidth]{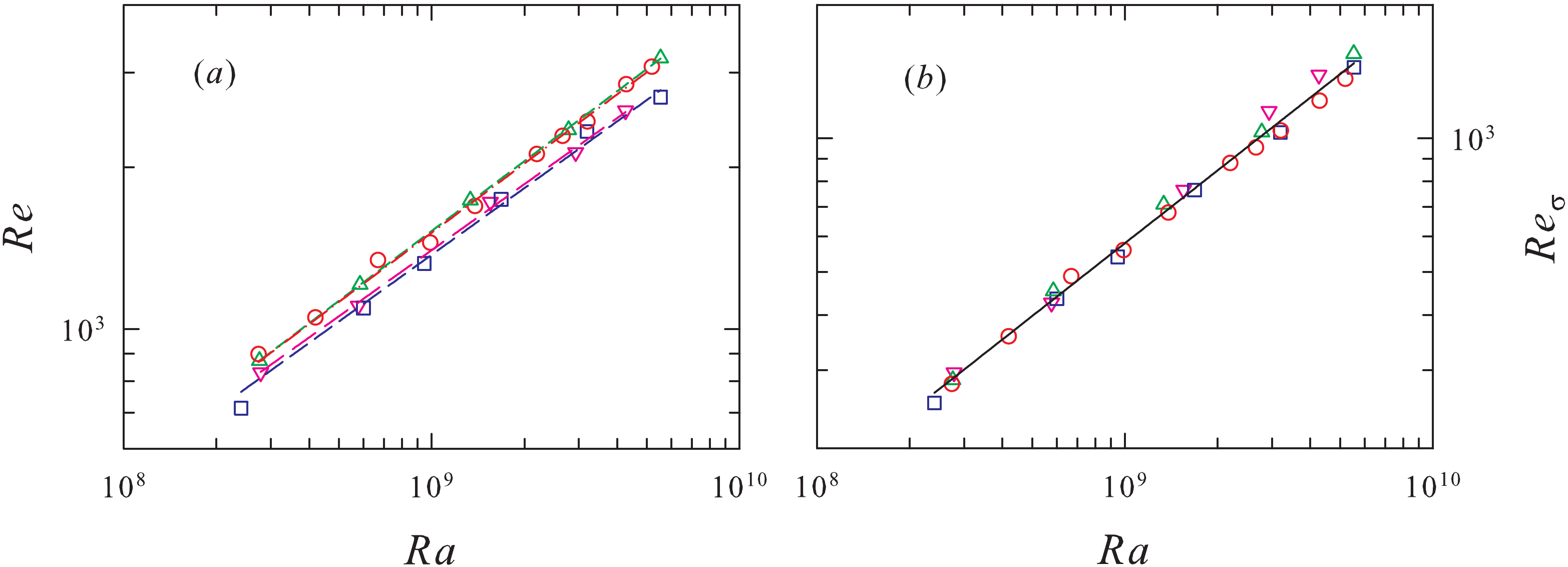}
}
\caption{\label{fig:ReandReRMS} (a) $Re$ based on the maximum horizontal velocity $U_{max}$,  and (b) $Re_{\sigma}$ based on the maximum velocity fluctuation $\sigma_{max}$ as a function of $Ra$ for different tilting angles. Inverted triangles: $\theta=0.5^{o}$; squares: $1^{o}$; triangles: $2^{o}$; circles: $3.4^{o}$. The dashed lines in (a) represent power-law fits to the individual data sets all with a scaling exponent $-0.43$ (see text for the fitting results). The solid line in (b) is a power law fit to all data sets in the plot, which gives $Re_{\sigma}=0.007 Ra^{0.55\pm0.01}$.}
\end{figure}

Figure~\ref{fig:AveragedField} (a) shows an example of measured instantaneous velocity map and (b) time-averaged velocity field taken over a period of $3.5$ h (corresponding to $25200$ velocity frames), with the cell tilted at $\theta=3.4^{o}$ and at $Ra=4.2\times10^{8}$. In the present measurement, $x$ spans from $-8.75$ mm to $8.75$ mm, and $z$ spans from $0$ to $17.5$ mm. From the velocity scale in figure ~\ref{fig:AveragedField} (a) and (b), it is seen that there exist velocity bursts with values much larger than the maximum velocity in the time-averaged velocity field. It is found that velocity maps measured at other tilt angles have similar features. As the mean velocity and the velocity fluctuations do not exhibit any obvious dependence on the horizontal position $x$ over the small range of the measurement, the quantities presented below are based on values averaged along the $x$-direction over the width of the measuring area. 

Figure~\ref{fig:UnNormalizedU} plots the velocity profiles for different tilt angles and various values of $Ra$, which shows that the shapes of the profiles are rather similar at this level of detail. Figure~\ref{fig:NormalizedU} plots normalized profiles in which $U(z)$ is normalized by the maximum horizontal velocity $U_{max}(Ra)$ (for ease of reference the values of $U_{max}$ are also listed in Table~\ref{table:parameters}) and the distance $z$ from the wall by the viscous boundary layer thickness $\delta_{v}(Ra)$ (to be defined below). The figure shows that up to $2\delta_{v}$ profiles for different $Ra$ and for the same tilt angle collapse on to a single curve quite well (except perhaps those correspond to the largest $Ra$ for $\theta =2.0$ and $3.4^{o}$).  Note that $z \simeq 2\delta_{v}$ is around where $U$ reaches its maximum value and beyond this position it decays toward cell centre. So this position may be taken as the separation between the boundary layer region and the bulk. The above results suggest that for the same tilt angle the profiles  in the boundary layer region have an invariant shape with respect to different values of $Ra$. This result is consistent with the finding by \citet{sun2008jfm}. In figure~\ref{fig:NormalizedU} we also plot the  theoretical Prandtl-Blasius profile. It is seen that within the BL ($z \leq \delta_{v}$) the profiles match the theoretical solution very well, while in the region just outside the boundary layer where plume emissions occur, all measured profiles are generally less steep than the Prandtl-Blasius profile. This feature is also similar to that observed by \citep{zhou2010prl} and will be further discussed in Sec.~\ref{subsec:Instantaneous boundary layer properties}. On the other hand, it is seen from figure~\ref{fig:NormalizedU} that profiles obtained at different $\theta$ seem to have different degrees of deviation from the Prandtl-Blasius profile. This can be seen more clearly in figure~\ref{fig:Rescale-TwoRa} where we show two examples in which profiles for different $\theta$ but with values of $Ra$ close to each other are plotted together along with the theoretical PB profile. This result suggests that the shape of the velocity profile near the plume-emission region is modified by the tilting angle. It is also noted that the profiles measured with $\theta = 1.0^{o}$ show strong deviations from the linear dependence with zero interception. We shall come back to this when discussing boundary layer scalings in the next section. 

Taking $U_{max}$ as the characteristic velocity of LSC, we define the Reynolds number $Re=U_{max}H/\nu$ and plot $Re$ as a function of $Ra$ and for different $\theta$ in figure ~\ref{fig:ReandReRMS}(a).  When fitting a power-law to the data for different $\theta$ separately, they all produce an exponent close to 0.43. To better compare the amplitude of $Re$ for different $\theta$, we fix the scaling exponent at $0.43$ and fit power laws to the different data sets again. This gives $Re=(0.185\pm0.002, 0.182\pm0.003,0.206\pm0.001,0.203\pm0.001)\times Ra^{0.43}$, where the amplitudes in the brackets are for $\theta=0.5^{o}$, $1.0^{o}$, $2.0^{o}$, and $3.4^{o}$, respectively. These results show that in general the values of $Re$ with larger $\theta$ are larger than those with smaller $\theta$. In an earlier study of the effect of cell titling, \citet{ahlers2006jfm} have found that $Re$ obtained indirectly from temperature measurement increases with the tilted angle, which is consistent with the trend observed here. We note also that the value of the scaling exponent of $Re$ obtained from many previous studies, and sometimes under nominally similar conditions, varies over a rather wide range from $0.43$ to $0.55$ (see for example, \citet{xin1996prl,xin1997pre,qiu1998pre_b,qiu1998pre,steinberg1999prl,lam2002pre,brown2007jsm,sun2008jfm,xie2012jfm}). The reason for such variations is not completely clear at present. A detailed study on this issue is beyond the scope of this paper. For interested readers, we refer to \citet{sun2005pre_b}  who offered an explanation that can account some of these dispersions in the exponent.

From the measured profile of the RMS velocity (see figure ~\ref{fig:ToFindThickness}), we can define another Reynolds number $Re_{\sigma}=\sigma_{max}H/\nu$, which is shown in figure ~\ref{fig:ReandReRMS}(b) as a function of Ra in a log-log scale for the four tilt angles. Here it is seen that $Re_{\sigma}$ does not seem to have an obvious dependence on $\theta$. We therefore fitted a single power law to all four data sets on the plot, which gave
$Re_{\sigma}=0.007 Ra^{0.55\pm0.01}$.  The value of the exponent is somewhat larger than $0.5$ that was obtained from several previous studies
 \citep{xin1996prl,xin1997pre,qiu1998pre_b,qiu1998pre,sun2008jfm}. But given the uncertainties in the experimental measurements, it is hard for one to attach too much significance to this difference.

\subsection{The viscous boundary layer and its scaling with $Ra$ and $Re$}\label{subsec:Scaling for Ra-scan}

\begin{figure}
\centerline{
\includegraphics[width=0.85\textwidth]{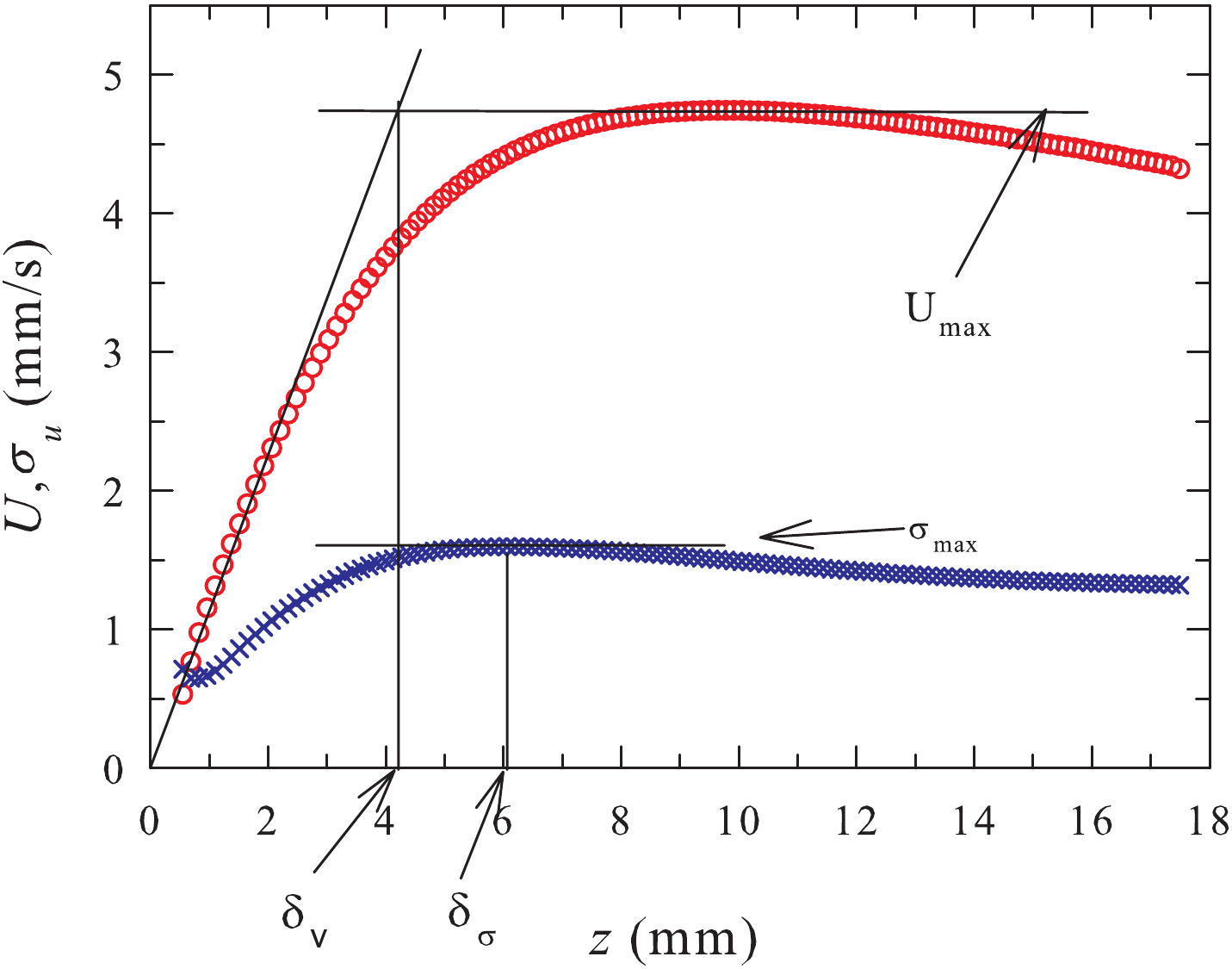}
}
\caption{\label{fig:ToFindThickness} Determination of the viscous boundary layer thickness $\delta_{v}$ through the slope-method from the mean horizontal velocity profile $U(z)$ (circles) and the thickness $\delta_{\sigma}$ from the standard deviation profile $\sigma_{U}$ (crosses). The measurement was made near the bottom plate with tilt angle $\theta=3.4^{o}$ and at $Ra=4.2\times10^{8}$.}
\end{figure}

We define the thickness $\delta_{v}$  of the viscous boundary layer through the ``slope-method" as shown in figure~\ref{fig:ToFindThickness} where a mean velocity $U(z)$ (circles) profile and the corresponding standard deviation profile $\sigma_{u}(z)$ (crosses) are shown, which are measured at $Ra=4.2\times10^{8}$ with $\theta=3.4^{o}$.   It is seen that $\delta_{v}$  is defined as the distance at which the extrapolation of the linear part of $U(z)$ equals its maximum value $U_{max}$, i.e. $\delta_{v}=U_{max}[dU/dz|_{z=0}]^{-1}$.  A length scale $\delta_{\sigma}$  can also be defined from the profile of $\sigma_{u}(z)$ where $\sigma_{u}$ reaches its maximum value. For the present example, the values for the two boundary layer length scales $\delta_{v}$ and $\delta_{\sigma}$ are found to be $4.20$ and $6.05$  mm, respectively. For ease of reference, the values of $\delta_{v}$ are listed in Table~\ref{table:parameters}. 

\begin{figure}
\centerline{
\includegraphics[width=0.7\textwidth]{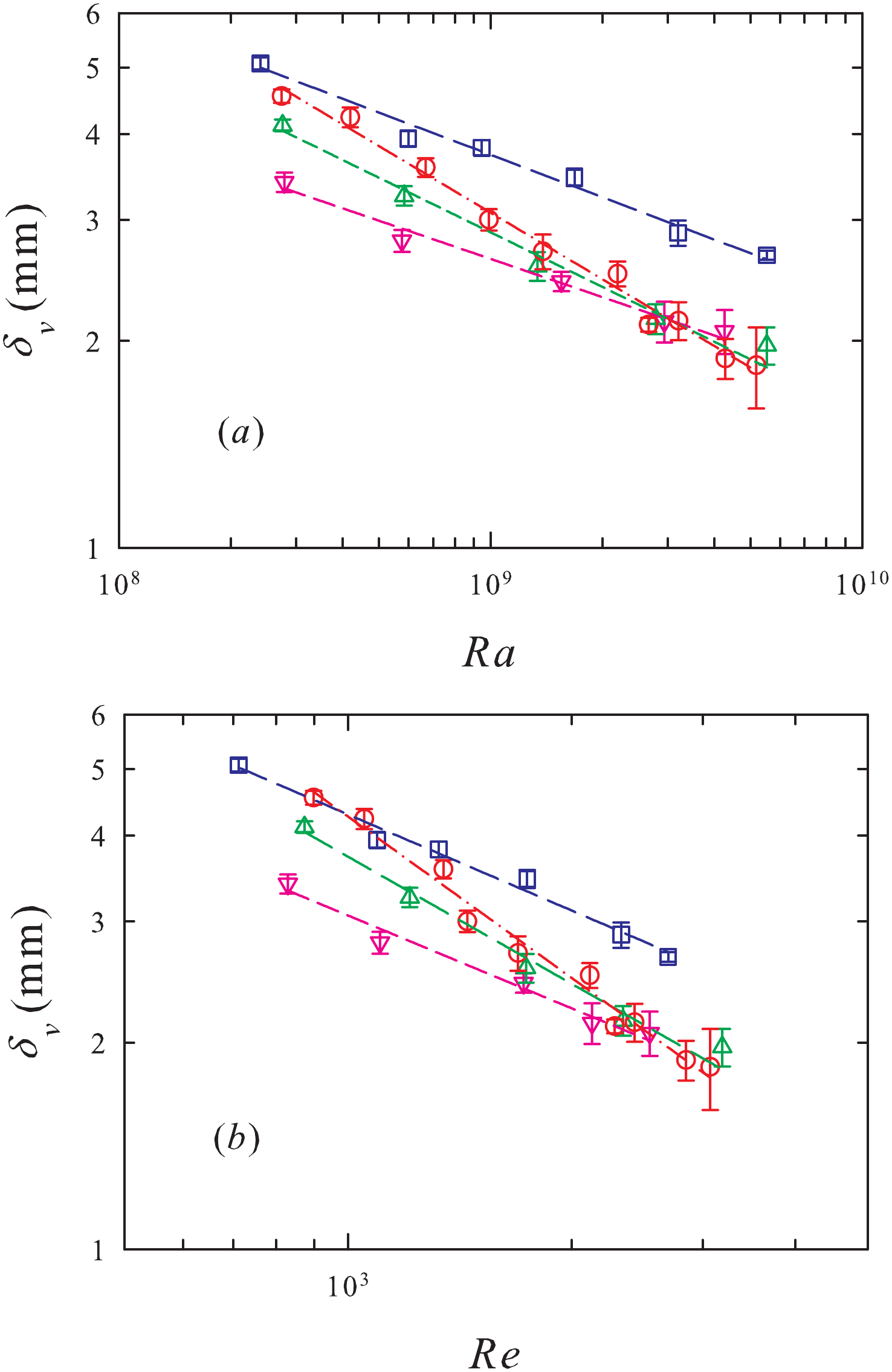}
}
\caption{\label{fig:thicknessv} Measured viscous boundary layer thickness  $\delta_{v}$  (a)versus $Ra$  and  (b) versus $Re$ for the four tilt angles: $\theta=0.5^{o}$ (inverted triangles); $1.0^{o}$ (squares); $2.0^{o}$ (triangles); and $3.4^{o}$ (circles). The dashed lines are power-law fits  $\delta_{v}/H =A_{1}Ra^{\beta_{1}}$ and $\delta_{v}/H =A_{2}Re^{\beta_{2}}$  to the respective data sets, with the fitting results listed in Table~\ref{tab:table2}.}
\end{figure}

\begin{figure}
\centerline{
\includegraphics[width=0.7\textwidth]{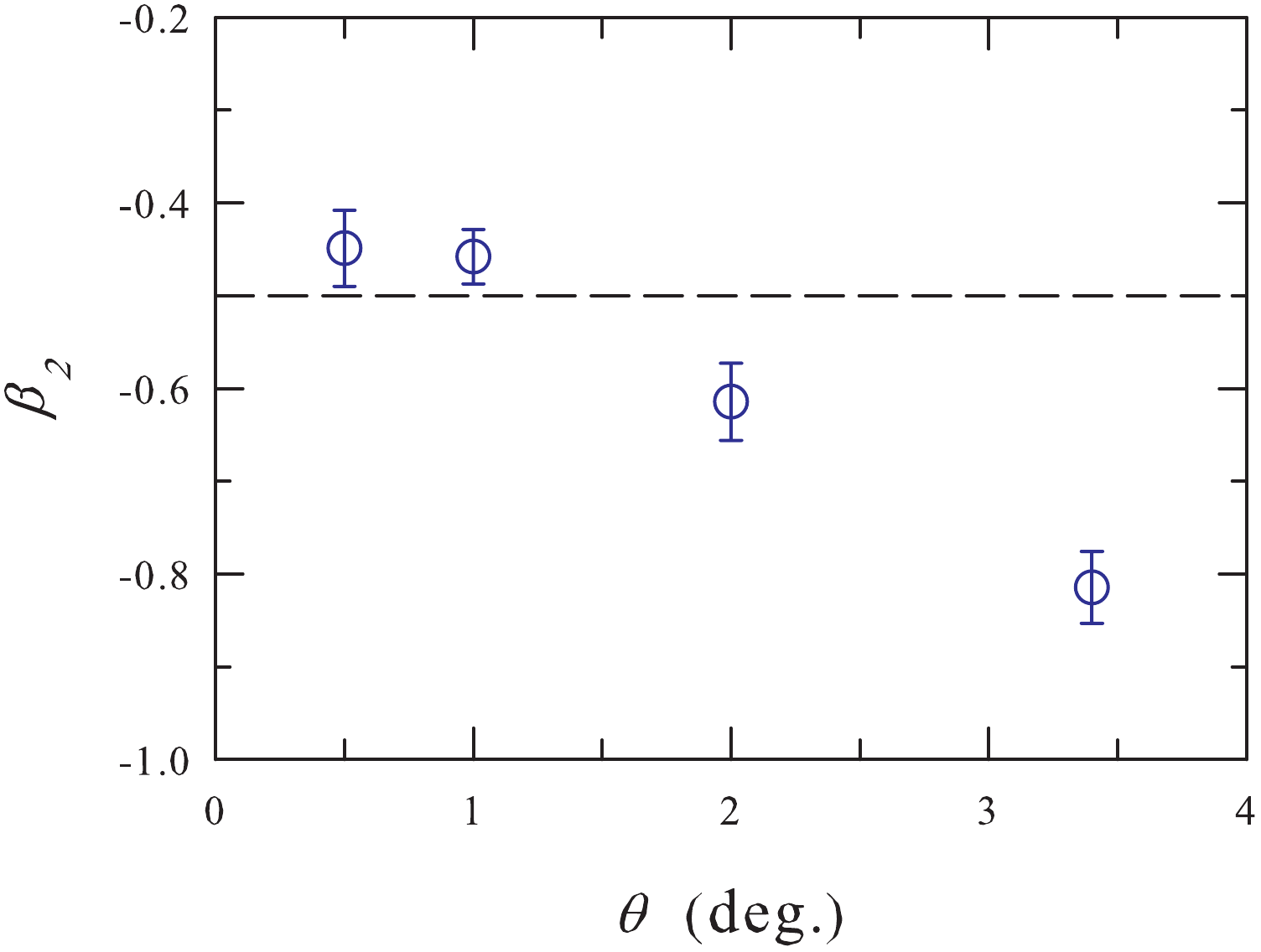}
}
\caption{The scaling exponent $\beta_{2}$ versus cell tilt angle $\theta$, where $\beta_{2}$ is obtained from the power law fit $\delta_{v}/H \sim Re^{\beta_{2}}$. The dashed line indicates $\beta_{2}=-0.5$ for a Pradtl-Blasius laminar boundary layer.
}
\label{fig:WithAngle} 
\end{figure}

We now examine the scalings of the boundary layer thickness with both the Rayleigh number $Ra$ and the Reynolds number $Re$.  In figures~\ref{fig:thicknessv}(a) and (b) we plot the measured viscous boundary layer thickness $\delta_{v}$ vs, respectively, $Ra$ and $Re$  for the four tilt angles. The lines in the figures represent the best power-law fits  $\delta_{v}/H =A_{1}Ra^{\beta_{1}}$ and $\delta_{v}/H =A_{2}Re^{\beta_{2}}$ to the respective data sets and the obtained fitting parameters are listed in Table~\ref{tab:table2}. Also shown in the Table for comparison are results obtained in cells with different geometries and using different methods. It is seen from the table that for small tilt angles ($\theta=0.5$ and $1.0^{o}$), the exponents are essentially the same and within the experimental uncertainties the $Re$-scaling exponent may be taken as the same as that predicted for a Prandtl-Blasius boundary layer, i.e. $\delta_{v} \sim Re^{-1/2}$. For larger titling angles, there appears  to be a trend for both $\beta_{1}$ and $\beta_{2}$ to decrease (absolute value increases) with increasing $\theta$.  It thus appears that titling the cell by over $1^{o}$ is a rather strong perturbation to the BL, at least for its scaling. The situation for the amplitude of viscous boundary layer thickness $\delta_{v}$ is a bit more complicated. From both figures~\ref{fig:thicknessv}(a) and (b) it seems that at lower values of $Ra$ ($Re$) the BL thickness increases with increasing tilting angle, except for $\theta = 1^{o}$. For this latter titling angle, $\delta_{v}$ appears to have an overall upward shift from the rest data sets.  While we do not know the exact reason(s) for this, we note from figure~\ref{fig:Rescale-TwoRa} that the profiles for this tilt angle seem to have a nonzero intercept on the horizontal axis. This appears to suggest that the origin of the $z$-axis for this was somehow shifted. But even if this is the case, the relatively small ``shift" cannot account for the large ``deviation" of this $\delta_{v}$ from the rest data sets (assuming there is indeed something ``wrong'' with this data set). Aside from the amplitude, the behavior of the $Ra$- and $Re$-scaling exponents may be summarized as follows. For small tilting angle ($\theta \le 1^{o}$), the effect of tilting is to lock the azimuthal plane of the LSC (or restrict its azimuthal meandering range) but the BL is otherwise not strongly perturbed and scaling wise the BL is approximately Prandtl-Blasius type. For relatively large titling angle  ($\theta > 1^{o}$), the BL appears to be strongly perturbed as far as scaling is concerned and the magnitude of the scaling exponent increases with titling angle, i.e. the BL thickness $\delta_{v}$ decays with increasing $Ra$ ($Re$) with a steeper slope. The situation is illustrated in figure~\ref{fig:WithAngle} where $\beta_{2}$ is plotted as a function of the tilt angle $\theta$.

%\begin{sidewaystable}
\begin{table}%The best place to locate the table environment is directly after its first reference in text
\begin{center}
\def~{\hphantom{0}}
\begin{tabular}{cccccccccc}
\textrm{Quantity}	&	\textrm{$Ra$ }	&	\textrm{Pr}	&	\textrm{Geometry}	&	\textrm{$A_{1}$}	&	\textrm{$-\beta_{1}$}	&	\textrm{$A_{2}$}	&	\textrm{$-\beta_{2}$}	&	$\theta (^{0})$	&	\textrm{Source}	\\
	&		&		&		&		&		&		&		&		&		\\
$\delta_{v}/H$	&	$10^{8}\sim 10^{10}$	&	$\sim 7$	&	cylin.	&	0.51	&	$0.16\pm0.02$	&		&	$0.32$	&	0	&	a	\\
	&	$10^{8}\sim10^{10}$	&	$6\sim1027$	&	cylin.	&	$0.65Pr^{0.24}$	&	$0.16\pm0.02$	&		&	$0.32$	&	0	&	b	\\
	&	$10^{8}\sim 10^{10}$ 	&	$\sim 7$	&	cubic	&	0.69	&	$0.18\pm0.04$	&		&	$0.36$	&	0	&	c	\\
	&		&		&		&	3.6	&	$0.26\pm0.03$	&		&	$0.52$	&	0	&	d	\\
	&	$10^{9}\sim 10^{10}$	&	$4.3$	&	rectan.	&	4.95	&	$0.27\pm0.01$	&	$0.64$	&	$0.50$	&	0	&	e	\\
	&	$10^{8}\sim 10^{9}$ 	&	$5.4$	&	cylin.	&	0.745	&	$0.19\pm0.01$	&	0.369	&	$0.45\pm0.04$	&	$0.5$	&	f	\\
	&		&		&		&	1.41	&	$0.20\pm0.01$	&	0.564	&	$0.46\pm0.03$	&	$1.0$	&	f	\\
	&		&		&		&	5.86	&	$0.29\pm0.01$	&	1.41	&	$0.61\pm0.04$	&	$2.0$	&	f	\\
	&		&		&		&	13.3	&	$0.32\pm0.01$	&	6.26	&	$0.81\pm0.01$	&	$3.4$	&	f	\\
	&		&		&		&		&		&		&		&		&		\\
	&		&		&		&	\textrm{$A_{3}$}	&	\textrm{$-\beta_{3}$}	&	\textrm{$A_{4}$}	&	\textrm{$-\beta_{4}$}	&		&		\\
$\delta_{\sigma}/H$	&	$10^{7}\sim 10^{11}$	&	$\sim 7$	&	cylin.	&	1.02	&	$0.25\pm0.02$	&		&	$0.5$	&	0	&	a	\\
	&	$10^{8}\sim 10^{10}$ 	&	$\sim 7$	&	cubic	&	0.95	&	$0.25\pm0.04$	&		&	$0.5$	&	0	&	c	\\
	&		&		&		&	43	&	$0.38\pm0.03$	&		&	$1.0$	&	0	&	d	\\
	&	$10^{9}\sim 10^{10}$	&	$4.3$	&	rectan.	&	16.5	&	$0.37\pm0.10$	&	$0.69$	&	$0.72\pm0.14$	&	0	&	e	\\
	&	$10^{8}\sim 10^{9}$ 	&	$5.4$	&	cylin.	&	0.58	&	$0.15\pm0.02$	&	$0.14$	&	$0.26\pm0.03$	&	$0.5$	&	f	\\
	&		&		&		&	1.77	&	$0.20\pm0.02$	&	$0.27$	&	$0.37\pm0.04$	&	$1.0$	&	f	\\
	&		&		&		&	2.68	&	$0.23\pm0.02$	&	$0.32$	&	$0.41\pm0.04$	&	$2.0$	&	f	\\
	&		&		&		&	9.9	&	$0.29\pm0.02$	&	$0.75$	&	$0.54\pm0.04$	&	$3.4$	&	f	\\

\end{tabular}
\caption{
%Exponents of scaling, with $Ra$ and $Re$, for the viscous boundary layer thickness ($\delta_{v}$) and rms velocity-based $\delta_{\sigma}$ for different geometries. The sources are: a. \citet{xin1996prl}; b. \citet{lam2002pre}; c. \citet{qiu1998pre}; d. \citet{sun2005pre}; e. present work.
Fitting results for the normalized viscous boundary layer thickness $\delta_{v}/H$ determined from the mean horizontal velocity profile and $\delta_{\sigma}/H$ determined from the rms horizontal velocity profile. The fitting parameters $A_{i}$ and $\beta_{i}$ ($i=1, 2, 3, 4$) are defined through the  power laws: $\delta_{v}/H=A_{1}Ra^{\beta_{1}}$, $\delta_{v}/H=A_{2}Re^{\beta_{2}}$, $\delta_{\sigma}/H=A_{3}Ra^{\beta_{3}}$, and $\delta_{\sigma}/H=A_{4}Re_{\sigma}^{\beta_{4}}$. The control parameters $Ra$ and $Pr$ and cell geometry of measurements are also listed. Also shown in the table are results from some previous experiments. The sources are: a. \citet{xin1996prl}; b. \citet{lam2002pre}; c. \citet{qiu1998pre_b} (bottom); d. \citet{qiu1998pre} (side wall); e. \citet{sun2008jfm}; and f. present work. (Note: the cell tilt angle $\theta$ is indicated as $0$ when it was not mentioned in the respective papers and we assume the cell was nominally leveled in those cases.)
}
\label{tab:table2}
\end{center}
\end{table}
%\end{sidewaystable}

\begin{figure}
\centerline{
\includegraphics[width=\textwidth]{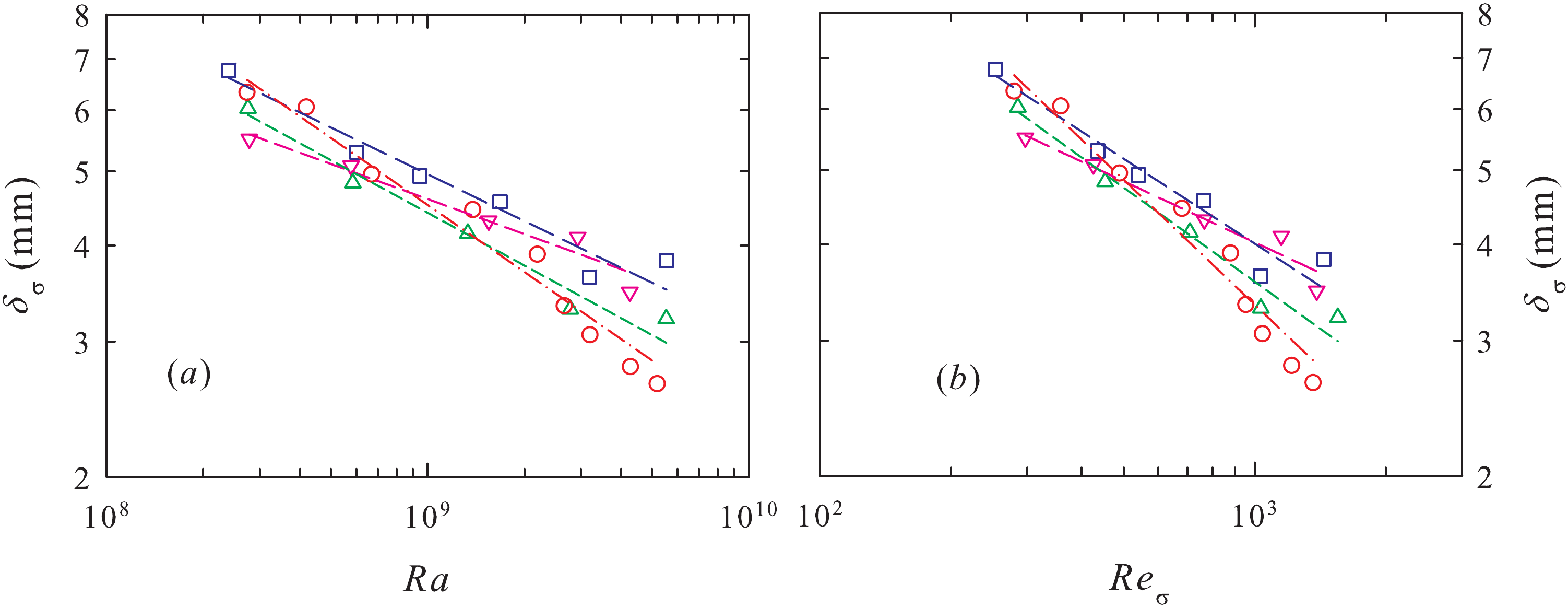}
}
\caption{\label{fig:delta-rms-VS-Re-rms} Scalings of the boundary layer scale $\delta_{\sigma}$ determined from the measured rms velocity profiles: (a) versus $Ra$  and (b) versus $Re_{\sigma}$. The symbols represent: $\theta=0.5^{o}$ (inverted triangles), $1.0^{o}$ (squares), $2.0^{o}$ (triangles), and $3.4^{o}$ (circles). The lines are power law fits $\delta_{\sigma}/H=A_{3}Ra^{\beta_{3}}$ and $\delta_{\sigma}/H=A_{4}Re_{\sigma}^{\beta_{4}}$ to the respective data sets, with the fitting results listed in Table~\ref{tab:table2}.}
\end{figure}

%We plot the exponents of thickness as a function of $Ra$ and $Re$ in figure~\ref{fig:WithAngle}. In the figure~
%\ref{fig:WithAngle}(a), obviously the error introduced in fitting is smaller than the difference between $\beta$ in different angle $\theta$, the $\beta$ is located between $-0.19$ and $-0.32$. In the figure~
%\ref{fig:WithAngle}(b), the $Re-$scaling exponents $\beta_{v-Re}$ of the velocity boundary layer span from $-0.45$ to $-0.81$. Before comparing with other result, we should understand and attempt to explain the change with tilted angle firstly.

In addition to the boundary layer thickness $\delta_{v}$  determined from the mean
horizontal velocity profile, another length scale can also be defined based on the profile
of the horizontal r.m.s. velocity $\sigma_{u}$, which may be called the r.m.s. velocity boundary layer thickness, as defined in figure~\ref{fig:ToFindThickness}. In figures~\ref{fig:delta-rms-VS-Re-rms}(a) and (b) we plot $\delta_{\sigma}$ versus $Ra$ and $Re_{\sigma}$ respectively. The $Ra$-scaling exponent varies from $-0.15$ to $-0.29$, which appears to follow similar trend as that of $\delta_{v}$, i.e. its absolute value increases with increasing $\theta$. But it and that of $Re_{\sigma}$-scaling exponent show significant difference with those obtained in previous studies. Table~\ref{tab:table2} shows the fitting results of $\delta_{\sigma}=A_{3}Ra^{\beta_{3}}$ and $\delta_{\sigma}=A_{4}Re_{\sigma}^{\beta_{4}}$.

Now we compare our result with previous experimental results obtained in the cells with different geometries. As shown in Table~\ref{tab:table2}, the value of $\beta_{1}$ obtained in both cylindrical and cubic geometries and measured near the bottom plate of the cell is $-0.16$. In all these previous measurements, the Reynolds number based on the maximum horizontal velocity near the plate was also obtained and they gave a scaling exponent $\gamma =0.5$ via $Re \sim Ra^{\gamma}$. From this we obtain $\beta_{2} = -0.32$.  In these studies, the convection cell was nominally leveled, i.e. not intensionally tilted. In the present study, for the   small tilting angle cases, where we assume the BL is not strongly perturbed, the measured $\beta_{1}  \simeq -0.19$ when combined with combined with $\gamma =0.43$ give a $\beta_{2} \simeq -0.45\pm 0.04$ (note that the actual value of $\beta_{2}$ are obtained from fitting the $\delta_{v}$ vs $Re$ data, not from  the relationship between the exponents). If we take these values to be close to the Prandtl-Blasius result, then scaling wise the viscous BL in a cylindrical geometry is also of a Prandtl-Blasius type, as was already found in a rectangular cell \citep{sun2008jfm}. For the relatively large deviations found in the untilted case, it may be attributed to the random azimuthal motion of the LSC.  

Finally we remark that as far as the scaling of the viscous BL is concerned, there is no theoretical prediction for the dependence of $\delta_{v}$ on $Ra$, only that on $Re$ (for example, $\delta_{v} \sim Re^{-1/2}$ for the Prandtl-Blasius BL). In the literature, it is sometimes stated that $\delta_{v}$ should scale as $Ra^{-1/4}$ for the Prandtl-Blasius BL. This is based on the assumption that $Re \sim Ra^{1/2}$. From above we have seen that the scaling exponent of $Re$ with $Ra$ varies over a rather wide range. It is therefore more meaningful to talk about the scaling of  $\delta_{v}$ with $Re$, rather than with $Ra$. We further note that in \citet{sun2008jfm} it was found that $\delta_{v} \sim Ra^{-0.27}$ and $Re \sim Ra^{0.55}$, which together give $\delta_{v} \sim Re^{-0.50}$. In the present case, we have $\delta_{v} \sim Ra^{-0.2}$ and $Re \sim Ra^{0.43}$, which together give $\delta_{v} \sim Re^{-0.46\pm0.03}$. Whether this is fortuitous or there is something deep here remains remains to be explored.
 
\subsection{Fluctuations and statistical properties of the velocity field in the boundary layer}\label{subsec:statistical properties of the velocity field in the boundary layer}

\begin{figure}
\centerline{
\includegraphics[width=\textwidth]{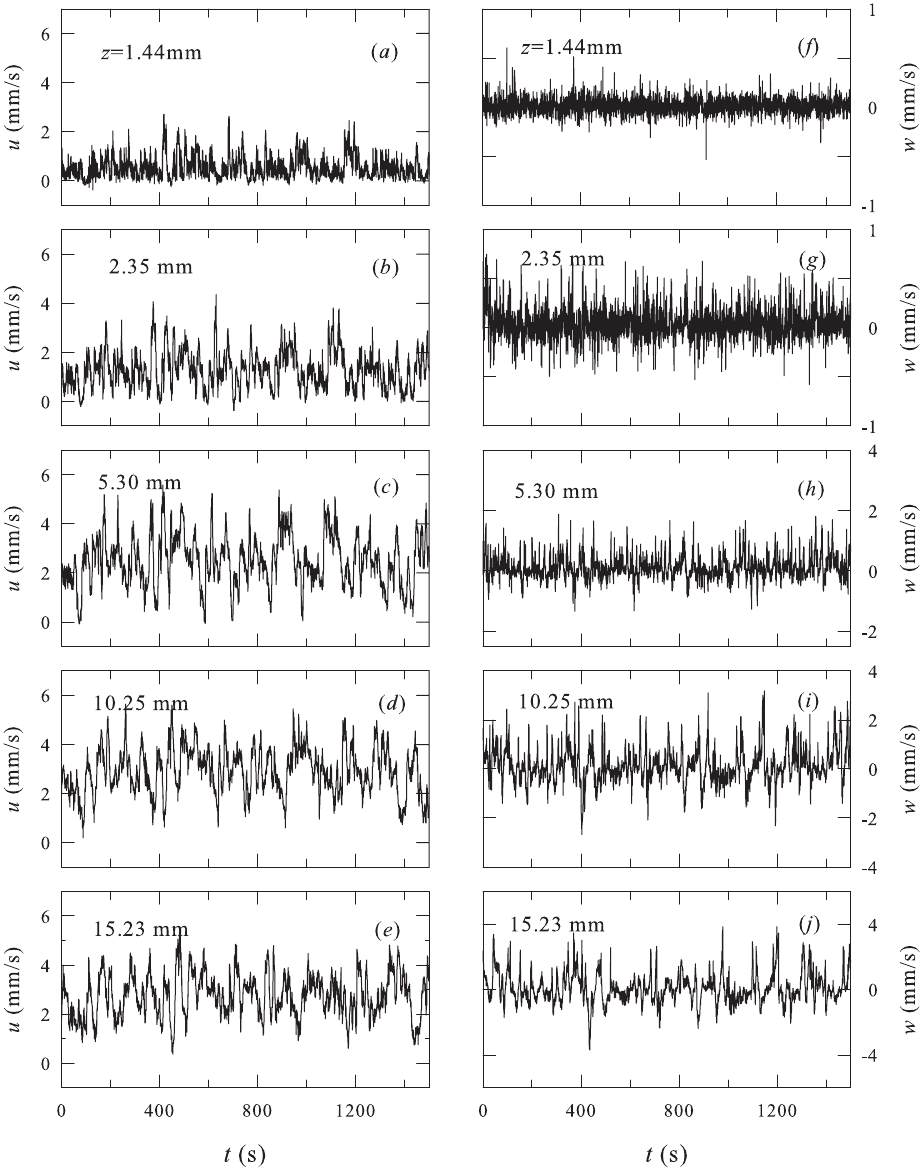}
}
\caption{Time traces of horizontal $u(t)$ (left panels) and vertical $w(t)$ (right panels) velocity components measured at $Ra=2.4\times10^{8}$ and $\theta=1^{o}$, at $x=0$ and different distances $z$ from the bottom plate.}
\label{fig:different}
\end{figure}

\begin{figure}
\centerline{
\includegraphics[width=\textwidth]{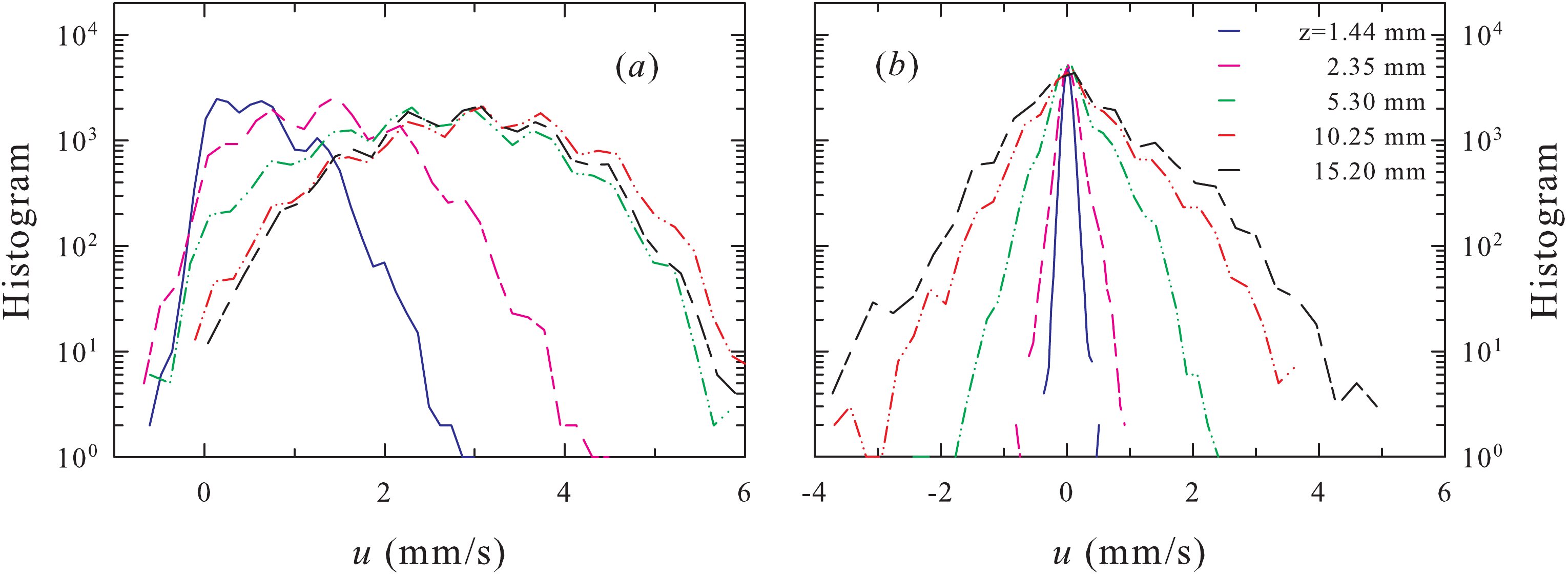}
}
\caption{Histograms of (a) the horizontal velocity $u(t)$ and (b) vertical velocity  $w(t)$  measured at various distances from the bottom plate with $Ra=2.4\times10^{8}$ and $\theta=1^{o}$.}
\label{fig:Different-Histogram}
\end{figure}

\begin{figure}
\centerline{
\includegraphics[width=\textwidth]{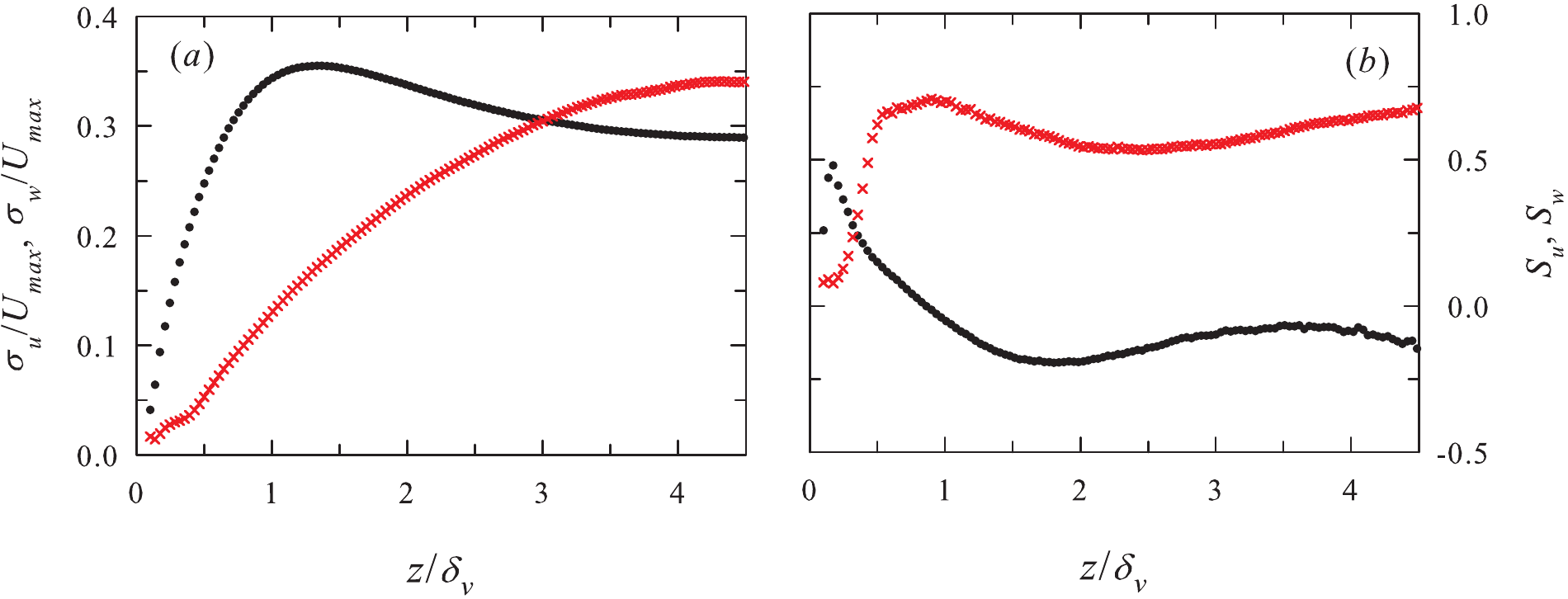}
}
\caption{Profiles of (a) the normalized rms velocity $\sigma_{u}$ ($\sigma_{w}$) and  of (b) the skewness $S_{u}$ ($S_{w}$) measured at $Ra=2.4\times10^{8}$ and $\theta=1^{o}$. The vertical distance $z$ is normalized by the velocity boundary layer thickness $\delta_{v}$. In both plots the circles represent those for the horizontal velocity component $u$ and the crosses represent those for the vertical velocity component $w$.}
\label{fig:rmsNormalized}
\end{figure}

In previous BL measurements in the cylindrical cell, owing to the nature of the dual-beam incoherent cross-correlation technique employed \citep{xin1996prl,lam2002pre}, only time-averaged velocity profiles are measured and no time-dependent quantities are obtained. It is therefore interesting to examine these quantities and compare them with similar quantities obtained in other type of turbulent flows. Figure~\ref{fig:different} shows the time series of both the horizontal component $u(t)$ (left panel) and the vertical component $w(t)$ (right panel) of the velocity, measured at various positions from the plate. The corresponding velocity histograms are shown in figure~\ref{fig:Different-Histogram}. The measurements were made at $Ra=2.4\times10^{8}$ and $\theta=1^{o}$. We show the velocity trace at several typical positions: (i) inside the thermal boundary layer, (ii) around the thermal boundary layer, (iii) around the viscous boundary layer; (iv) at the position of the maximum velocity; and (v) far away from the boundary layers. The figures show that the absolute horizontal velocity is much higher than vertical velocity at each position. One general feature we observed is that velocity time series and histograms look similar for different tilting angles. For this reason, we show here results for only one tilting angle. 

At $Ra=2.4\times10^{8}$, the viscous BL thickness is $\delta_{v}= 5.07$ mm. It is seen from figures~\ref{fig:different}(a) and ~\ref{fig:different}(b) that at positions inside the BL, the horizontal velocity $u(t)$ skews toward the positive side, i.e. the velocity is skewed toward the mean flow direction. This may be understood by the fact that close to the viscous sublayer the flow speed is very close to zero and a fluctuation smaller than the mean would mean a flow reversal, which is a rather rare event. Once outside of the BL, one observes more symmetric fluctuations around the mean velocity. For the vertical velocity $w(t)$, its mean velocity is very small at most positions. But the fluctuation increases significantly when the position is outside of the BL, which are signatures of plume emissions at these positions. These properties can also be seen from the velocity histograms shown in figure~\ref{fig:Different-Histogram}. A notable difference of the present results from those observed in \citet{sun2008jfm} is that for positions outside of the BL the horizontal velocity fluctuates more or less symmetrically around the mean, rather than skewed toward the negative as seen in the rectangular cell.  

The statistical properties of the velocity may be characterized more quantitatively by its root-mean-square (r.m.s.) value and its skewness, which are shown in figure~\ref{fig:rmsNormalized}. Figure~\ref{fig:rmsNormalized}(a) plots the velocity r.m.s $\sigma_{u}$ and  $\sigma_{w}$ normalized by maximum horizontal velocity $U_{max}$ versus the normalized distance $z/\delta_{v}$. Figure~\ref{fig:rmsNormalized}(b) shows the skewness profiles $S_{u}=\langle  (u-\langle u\rangle)^{3}\rangle/(\langle(u-\langle u\rangle)^{2} \rangle)^{3/2}$ and $S_{w}=\langle  (w-\langle w\rangle)^{3}\rangle/(\langle(w-\langle w\rangle)^{2} \rangle)^{3/2}$ for the horizontal and vertical velocities, respectively. 
Similar to \citet{sun2008jfm},  our result could not tell whether $\sigma_{w}$ favors a power law or a logarithmic scaling with the distance $z$, even though our measurement had a much higher spatial resolution. This is partly due to the limited size of the measurement area.

\subsection{Properties of shear stresses and near-wall quantities}\label{subsec:the quantities of turbulence}

\begin{figure}
\centerline{
\includegraphics[width=\textwidth]{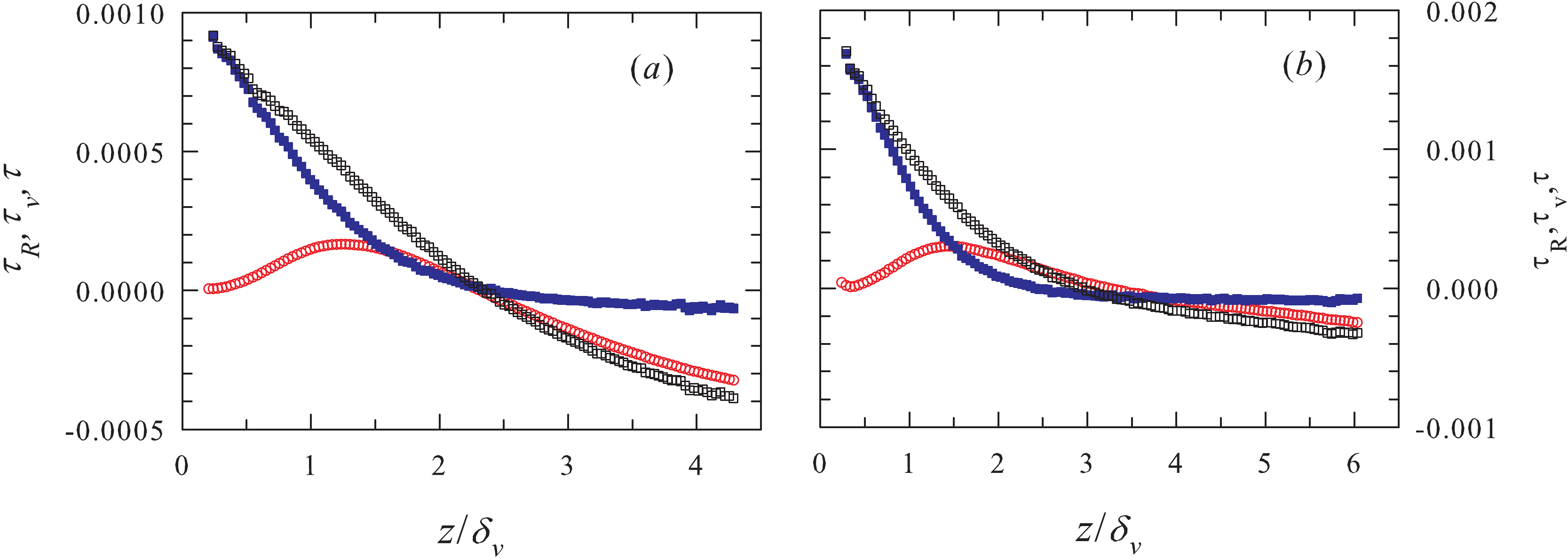}
}
\caption{\label{fig:ShearDistribution} Viscous stress $\tau_{v}$ (solid squares), the Reynolds stress $\tau_{R}$ (open circles) and the total stress $\tau$ (open squares) as functions of the normalized distance from the plate for (a) $Ra=4.2\times10^{8}$ and (b) $Ra=9.9\times10^{8}$, both with $\theta=3.4^{o}$.}
\end{figure}

\begin{figure}
\centerline{
\includegraphics[width=\textwidth]{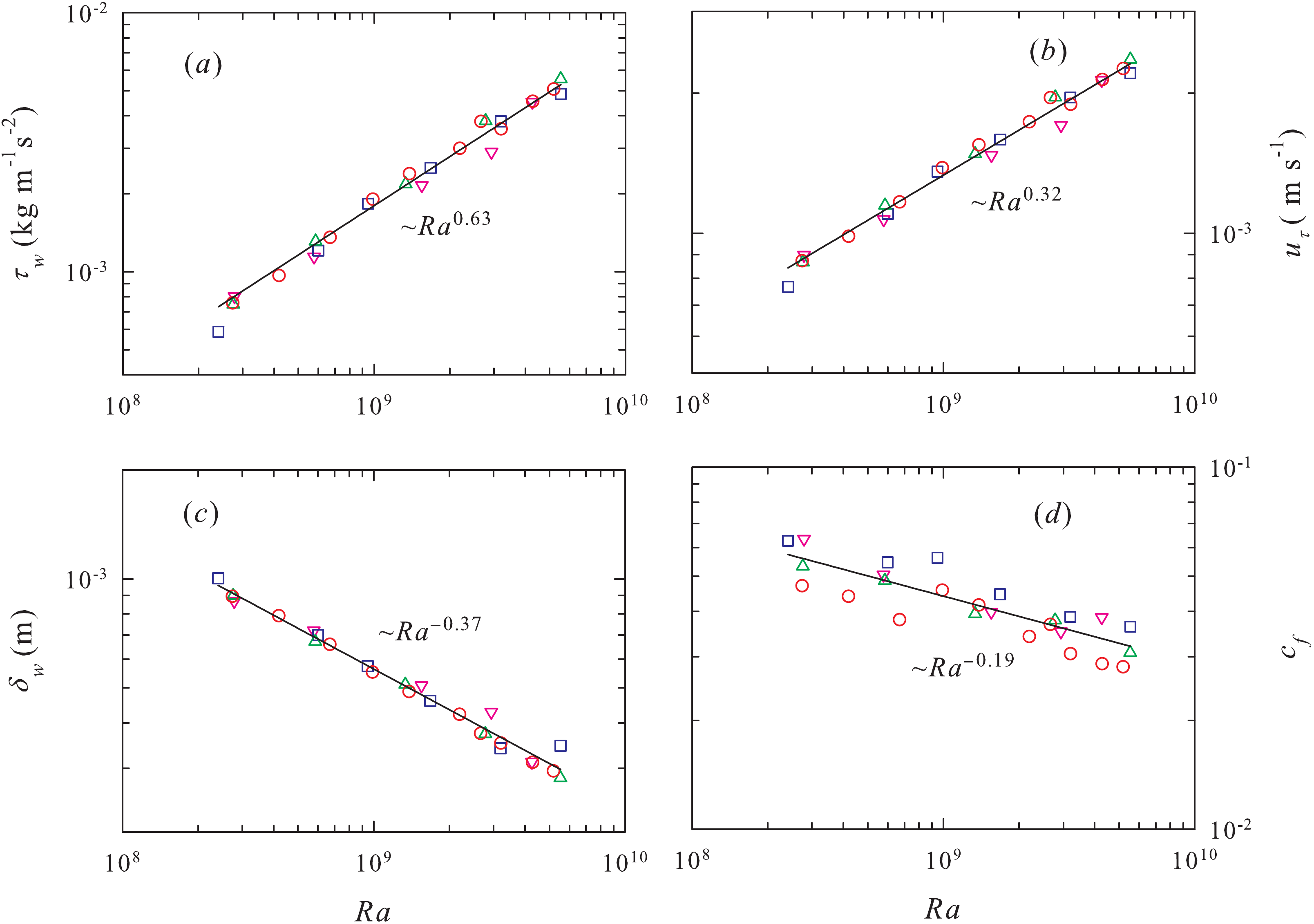}
}
\caption{\label{fig:wallRa} The \textit{Ra}-dependence of (\textit{a}) the wall shear stress $\tau_{w}$, (\textit{b}) the friction velocity $u_{\tau}$, (\textit{c}) the wall thickness (viscous sublayer) $\delta_{w}$, and (\textit{d}) the friction coefficient $c_{f}$, with power law fits shown as solid lines. The symbols represent data for different tilting angles: $\theta=0.5^{o}$ (inverted triangles), $1^{o}$ (squares), $2^{o}$ (triangles), and $3.4^{o}$ (circles).
}
\end{figure}

\begin{figure}
\centerline{
\includegraphics[width=\textwidth]{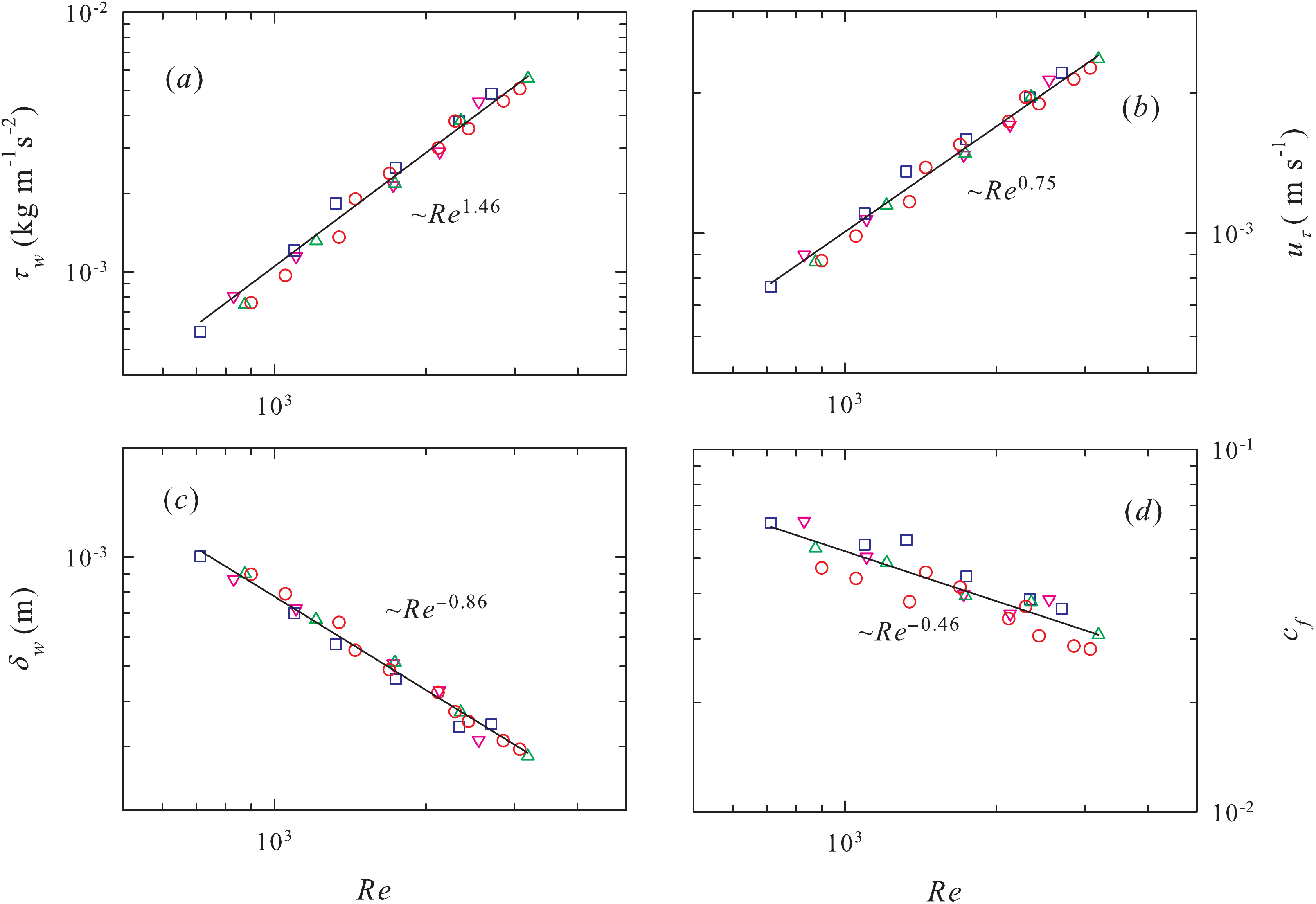}
}
\caption{\label{fig:wallRe} The \textit{Re}-dependence of (\textit{a}) the wall shear stress $\tau_{w}$, (\textit{b}) the friction velocity $u_{\tau}$, (\textit{c}) the wall thickness (viscous sublayer) $\delta_{w}$, and (\textit{d}) the friction coefficient $c_{f}$. The symbols represent data for different tilting angles: $\theta=0.5^{o}$ (inverted triangles), $1^{o}$ (squares), $2^{o}$ (triangles), and $3.4^{o}$ (circles). Power law fits are indicated in the figure.
}
\end{figure}
\begin{figure}
\centerline{
\includegraphics[width=0.75\textwidth]{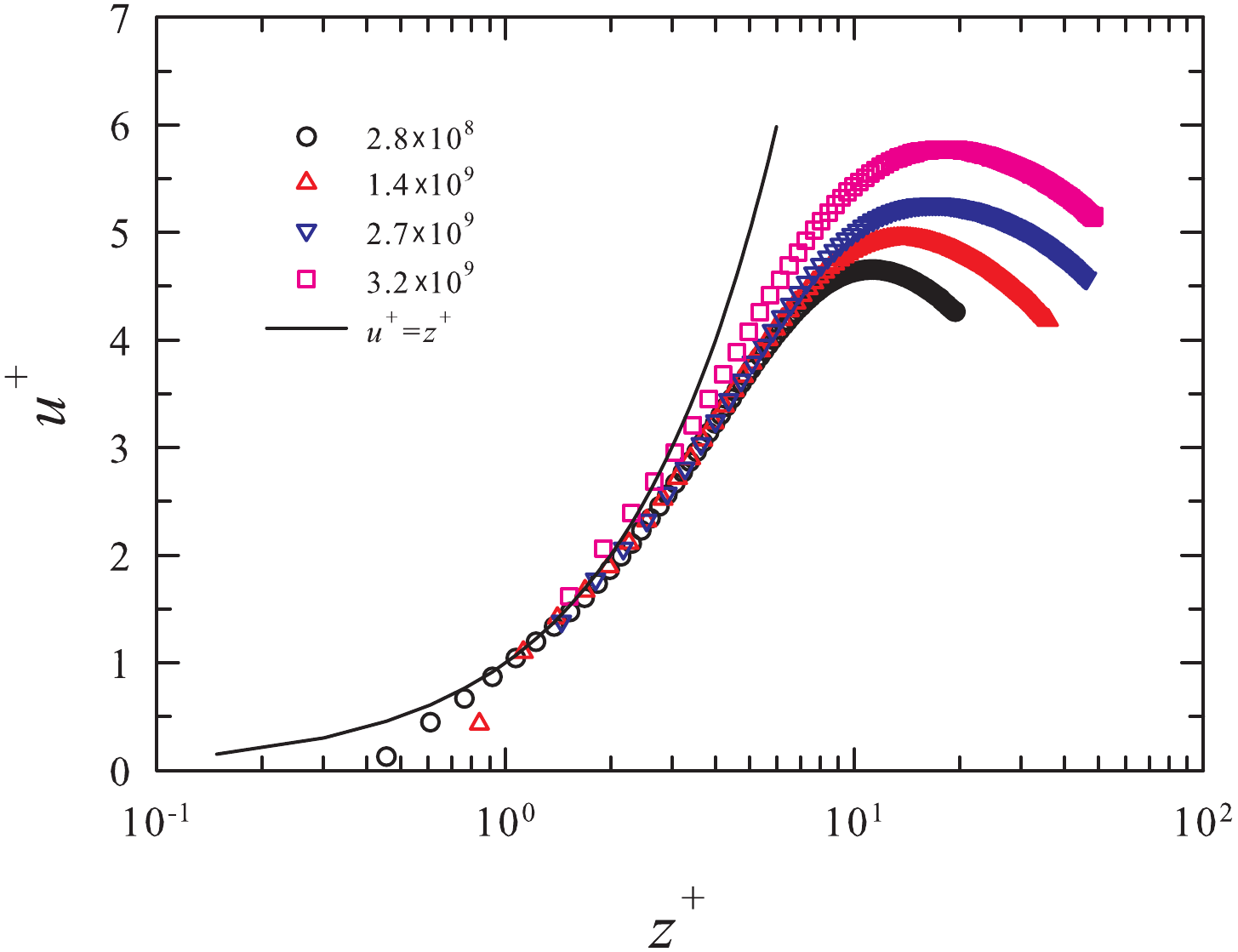}
}
\caption{\label{fig:wallNormalized} Measured horizontal velocity profiles normalized by wall units for four different Ra all taken at tilting angle $\theta=3.4^{o}$.}
\end{figure}

One of the advantages of PIV measurement is that it enables one to measure the horizontal and vertical velocities at the same time, so that one can calculate the Reynolds shear stress $\tau_{R}=-\rho(z)\langle u'w' \rangle$. Here $u'$ and $w'$ are the fluctuations of the horizontal and vertical velocity components respectively, $\rho(z)$ is the \textit{z-}dependent fluid density. Viscous shear stress is defined as $\tau_{v}=\mu(z)du/dz$, where $\mu(z)$ is the dynamic viscosity dependent on position $z$. The Reynolds stress  represents the transport of momentum by turbulent fluctuations, whereas the viscous stress describes the momentum transfer by viscosity. The total shear stress is then $\tau=\tau_{v}+\tau_{R}$.

Figure~\ref{fig:ShearDistribution} plots the profiles of the viscous shear stress, Reynolds stress and total stress for (a) $Ra=4.2\times10^{8}$, and (b) $9.9\times 10^{8}$. It is seen that both Ra have the same qualitative features. Here the examples are for $\beta = 3.4^{o}$, and results for other tilting angles are similar. 
Near the plate, it is seen that the Reynolds stress $\tau_{R}$ is close to zero, while the viscous shear stress $\tau_{v}$ is maximum because of the large velocity gradient $du/dz$ at the wall. So the total stress at the wall $\tau_{w} (=\tau(0))$ comes almost entirely from the contribution of the viscous shear stress. Moving away from the plate, the velocity gradient becomes smaller and the viscous shear stress decreases to zero.  The Reynolds stress $\tau_{R}$ increases and attains its maximum at $z\approx 1.5\delta_{v}$.  It then decreases to around zero $z\approx 2\delta_{v}$ and becomes negative in the bulk flow. It is also seen clearly from the figure that $\tau_{R}$ and $\tau_{v}$ cross at $z\approx 1.5\delta_{v}$, where $\tau_{R}$ is close to its maximum value. This suggests that the momentum transfer in the outer region is dominated by turbulent fluctuations. But in the viscous boundary layer, the momentum transfer is still dominated by the viscous diffusion, which implies that the viscous boundary layer  is still laminar in this range of $Ra$. 

%The universal laws of boundary layer in systems such as pipe flows have been well-studied theoretically, numerically and experimentally (Eggels1994,schlichting2000,Shang2002). 

With the measured near-wall high-resolution velocity field, we are now in a position to check the dynamic wall properties in turbulent thermal convection. We first consider the scaling of four basic wall quantities with both $Ra$ and $Re$. These are the wall shear stress $\tau_{w}$, the skin-friction velocity $u_{\tau}=(\tau/\rho_{0})^{1/2}$, the viscous sublayer length scale $\delta_{w}=\nu_{0}/u_{\tau}$, and the skin-friction coefficient $c_{f}=\tau/\rho_{0}U_{max}^{2}$. Here $\rho_{0}\equiv \rho(z=0)$ and $\nu_{0}\equiv \nu(z=0)$. Figure~\ref{fig:wallRa} shows the scaling of these quantities with $Ra$.  It is seen that within experimental uncertainties there is no difference between data with different $\theta$. This suggests that tilting the cell does not have any appreciable effect on BL properties near the wall. Without differentiating the different data sets, power law fits to all data yield $\tau_{w}\sim Ra^{0.63}$, $u_{\tau} \sim Ra^{0.32}$, $\delta_{w} \sim Ra^{-0.37}$ and $c_{f} \sim Ra^{-0.19}$. In a rectangular cell, Sun \textit{et al} (2008) found for the same quantities the fitted power law exponents $0.86$, $0.44$, $-0.50$, and $-0.28$ respectively. It is seen that the absolute values of these exponents are all larger than those obtained in the present experiment. There is no theoretical prediction for the \textit{Ra}-scaling of these quantities in turbulent thermal convection, so we do not know what the difference means.
 
It will be more useful perhaps to examine the scaling of these quantities with the Reynolds number $Re$, since theoretical predictions exist for such scalings for wall-bounded shear flows \citep{schlichting2000}.  Figure~\ref{fig:wallRe} plots these quantities as a function of \textit{Re}, the symbols are the same as in figure~\ref{fig:wallRa}. For the quantities $\tau_{w}$, $u_{\tau}$, $\delta_{w}$, and $c_{f}$ our results give the exponents $1.46$, $0.75$,  $-0.86$, and $-0.46$. 
For a laminar boundary layer over a flat plate, the theoretically predicted `classical' exponents for these quantities are $3/2$, $3/4$, $-1$, and $-1/2$ respectively. One sees that within the experimental uncertainties there is an excellent agreement between the present experiment and the theoretical predictions for all the wall quantities except for $\delta_{w}$, which is a bit smaller. 
For reference, the previous measurement in rectangular cell gives $1.55$, $0.8$, $-0.91$, and $-0.34$ for the corresponding quantities \citep{sun2008jfm}. 
%We caution, however, that given the short range of $Ra$ and the amount of data scatter (such as that seen in $c_{f}$), these results should be taken only as reference.

To further compare the present system with classical boundary layers, we examine velocity profiles in terms of the wall units. Figure~\ref{fig:wallNormalized} shows the normalized mean horizontal velocity profiles for four different values of $Ra$ taken at $\theta=3.4^{o}$ in a semi-log plot, here $u^{+}=u(z)/u_{\tau}$ and $z^{+}=z/\delta_{w}$. The linear scaling of $u^{+}$ over $z^{+}$ in the viscous sublayer below $z^{+}<5$ is reflected quite well by the measured profiles confirming that the boundary layer is not turbulent in the present range of $Ra$ and $Pr$. The velocity normalized by wall unit decrease after reaching the maximum value in $z^{+}\sim 10$. Comparing to the same quantity measured in the rectangular cell \citep{sun2008jfm}, however, our result shows some deviation from the theoretical profile. This is a reflection of the fact that in the cylindrical cell it is more difficult to measure the profile accurately very close to the wall.

\subsection{Dynamical scaling and the shape of velocity profiles in the boundary layer}\label{subsec:Instantaneous boundary layer properties}

\begin{figure}
\centerline{
\includegraphics[width=\textwidth]{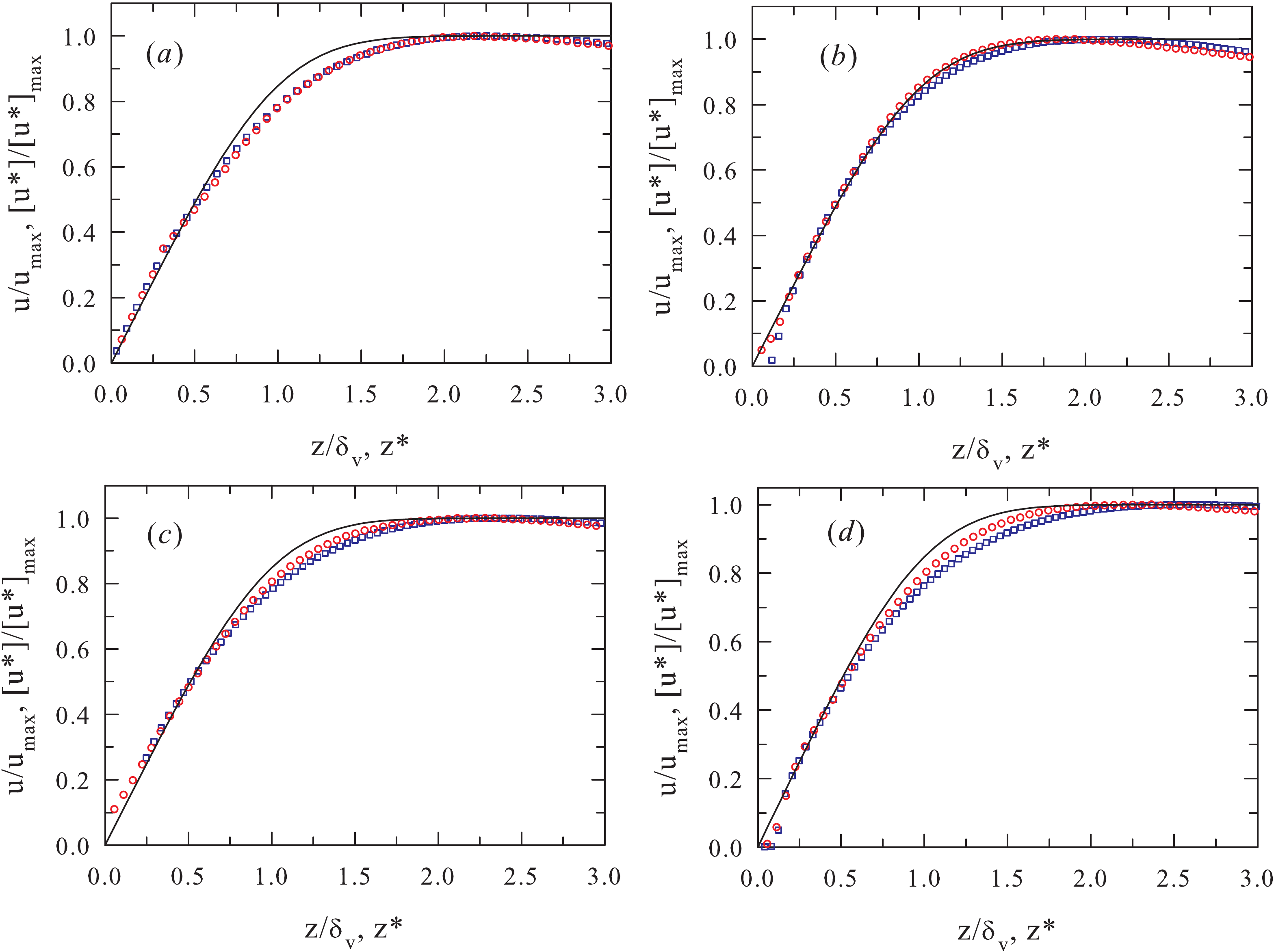}
}
\caption{ Comparison between profiles obtained in the dynamical frame ($u^{*}(z^{*})$, red circles) and the laboratory frame ($u(z)$, blue squares), measured at different tilt angles $\theta$ but with comparable values of $Ra$. Also shown for comparison is the theoretical Prandtl-Blasius laminar velocity profile (solid line). (a) $\theta=0.5^{o}$, $Ra=5.77\times10^{8}$; (b) $\theta=1.0^{o}$, $Ra=6.00\times10^{8}$; (c) $\theta=2.0^{o}$, $Ra=5.85\times10^{8}$; and (d) $\theta=3.4^{o}$, $Ra=6.69\times10^{8}$.}
\label{fig:Profile-6e+8}
\end{figure}

\begin{figure}
\centerline{
\includegraphics[width=\textwidth]{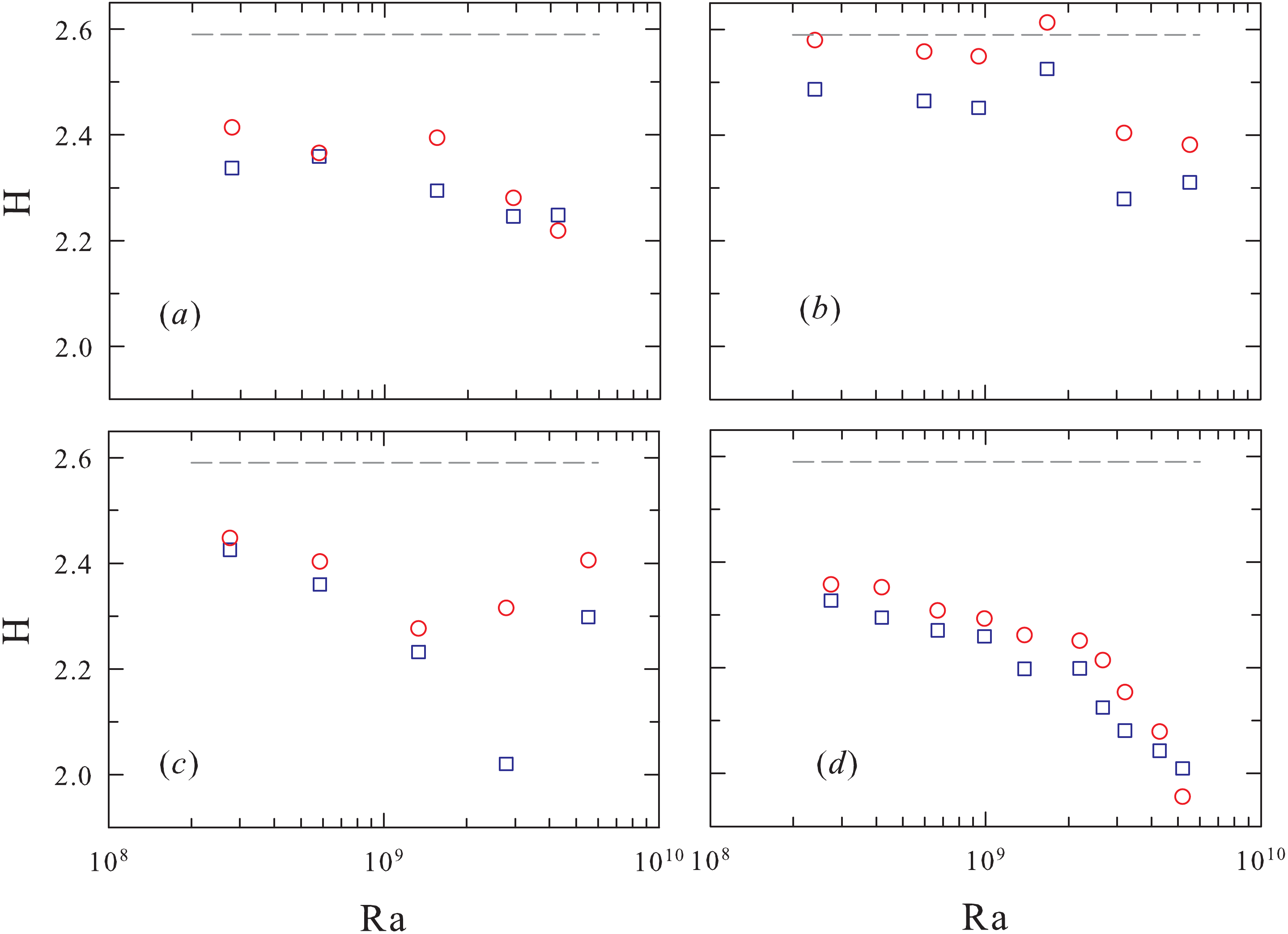}
}
\caption{ The shape factor $H=\delta_{d}/\delta_{m}$ of profiles $u^{*}(z^{*})$ (red circles) obtained in the dynamical frame and of $u(z)$ (blue squares) obtained in the laboratory frame as a function of $Ra$ and for different titling angles.  (a) $\theta=0.5^{o}$, (b) $1.0^{o}$, (c) $2.0^{o}$, and (d) $3.4^{o}$. The dashed line represents the value of $2.59$ for the theoretical Prandtl-Blasius laminar BL.}
\label{fig:H-four-angles}
\end{figure}

\begin{figure}
\centerline{\includegraphics[width=\textwidth]{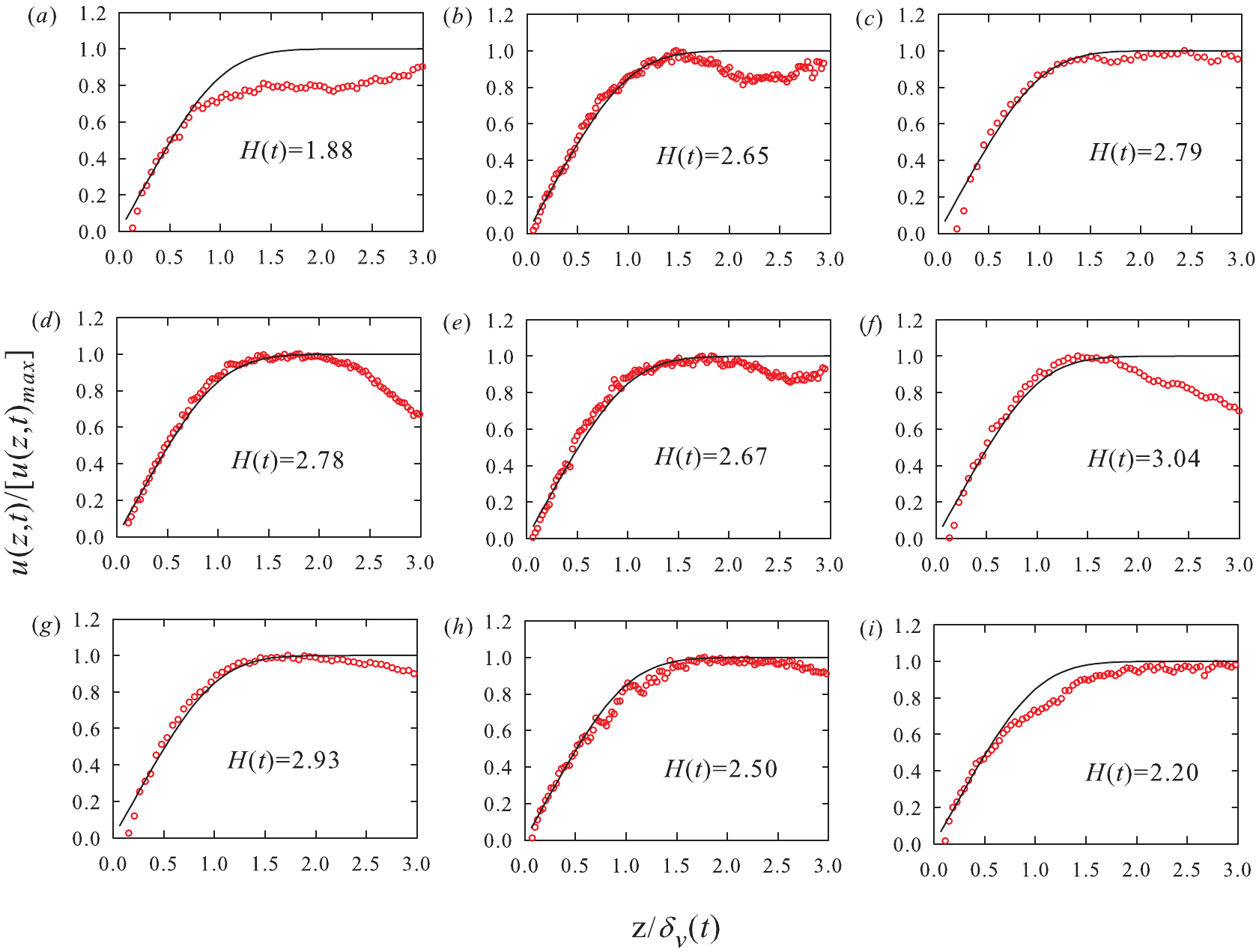}
}
\caption{ Examples of instantaneous horizontal velocity profiles scaled by the instantaneous kinematic BL thickness and the instantaneous maximum velocity (measured at $Ra=6.00\times10^{8}$ and $\theta=1^{o}$). The corresponding instantaneous shape factor is also indicated on the plot. The solid curves are the Prandtl-Blasius velocity profiles.}
\label{fig:1degree-6E+8-Instantaneous}
\end{figure}

\begin{figure}
\centerline{
\includegraphics[width=\textwidth]{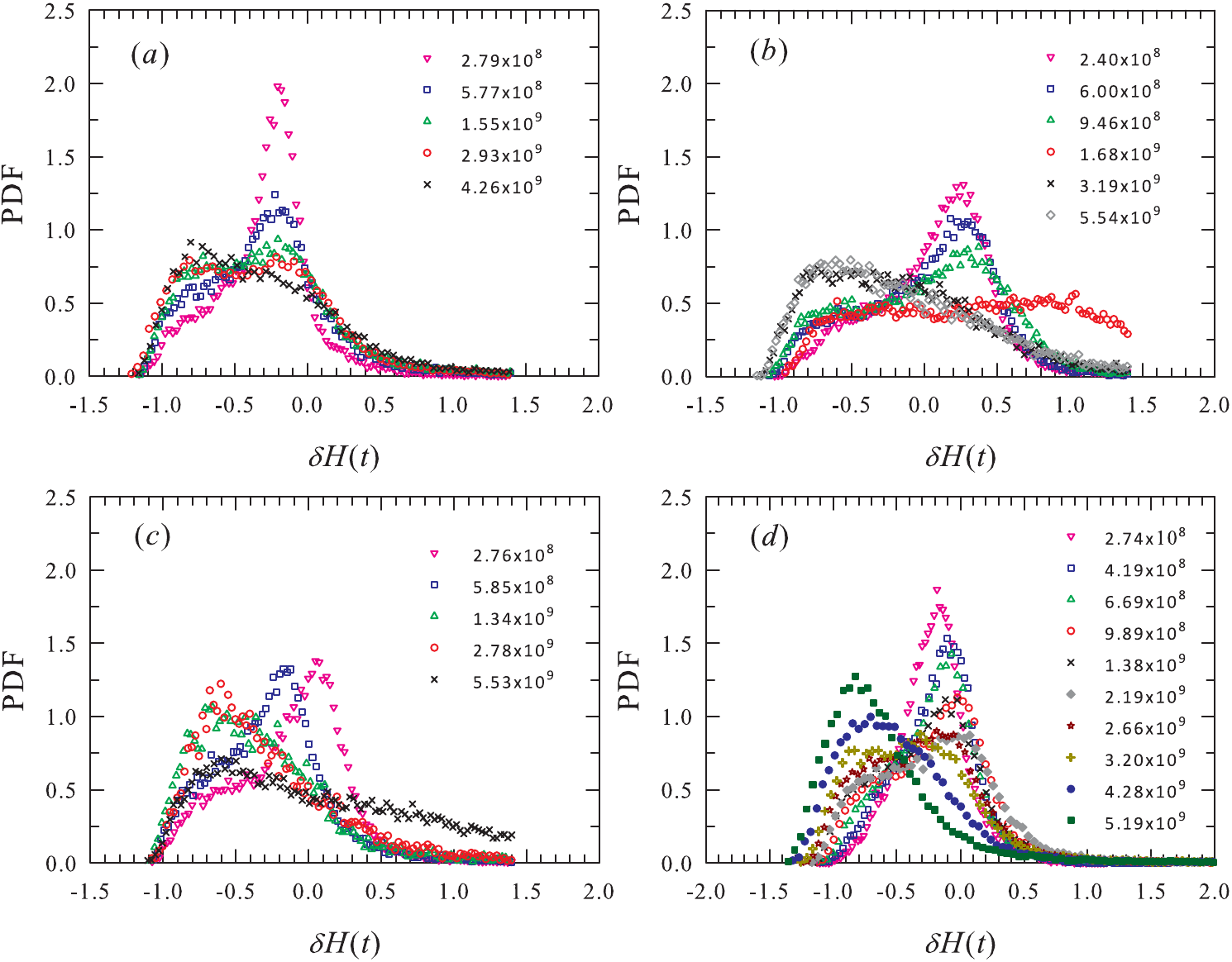}
}
\caption{ PDFs of the shape-factor difference $\delta H$ between those of the rescaled instantaneous profiles and that of the Prandtl-Blasius profile,  measured at (a) $\theta=0.5^{o}$, (b) $1.0^{o}$, (c) $2.0^{o}$, and (d) $3.4^{o}$.}
\label{fig:DeltaH-pdf}
\end{figure}

The dynamic scaling method of \citet*{zhou2010prl} has been found to work well when tested in quasi-2D experiment and 2D numerical simulations \citep{zhou2010jfm,zhou2011pof}. But it has not been examined in 3D experiments. Here we investigate how it works in our cylindrical geometry. As the method has been well documented elsewhere \citep{zhou2010prl,zhou2010jfm,zhou2011pof}, we will only give a brief description of it here. From the measured instantaneous velocity profile $u(z,t)$ one can obtain an instantaneous viscous boundary layer thickness $\delta_v(t)$ using the same `slope' method as used for the mean velocity profiles.
A local dynamical BL frame can then be constructed by defining the time-dependent rescaled distance
$z^*(t)$ from the plate as
\begin{equation}
z^*(t) \equiv z/\delta_v(t).
\end{equation}
The dynamically time averaged mean velocity  profile $u^*(z^*)$ in the dynamical BL frame is then obtained
by averaging over all values of $u(z,t)$ that were measured at different discrete times $t$ but at the same relative position $z^*$, i.e.
\begin{equation}
u^*(z^*) \equiv \langle u(z,t)|z = z^*\delta_v(t)\rangle.
\end{equation}

Figure ~\ref{fig:Profile-6e+8} shows the mean velocity profiles measured in the laboratory and the dynamical frames respectively at the four tilting angles and for comparable values of $Ra$ (as indicated in the figure caption). These results show that the dynamical scaling method appears to be more effective for larger values of $\theta$. This may be understood based on the fact that a larger tilt angle will place stronger restriction on the azimuthal meandering of the LSC so that it has less fluctuations in the horizontal direction perpendicular to the mean flow. We note, however, regardless of the tilt angle, the method works less effectively than it is in quasi-2D experiment and 2D simulations. 

A more quantitative approach to characterize the shape of the mean velocity profiles is to investigate their shape factor $H=\delta_{v}/\delta_{m}$ defined as the ratio between the displacement thickness $\delta_{v}$ and the momentum thickness $\delta_{m}$, where
\begin{equation}
\delta_{d}=\int_{0}^{\infty}[1-\frac{u(z)}{u_{max}}]dz, ~\protect{and}~ 
\delta_{m}=\int_{0}^{\infty}[1-\frac{u(z)}{u_{max}}]\frac{u(z)}{u_{max}}dz.
\end{equation}
Since $u(z)$ decays after reaching its maximum value, the above integrations are evaluated only over the range from $z=0$ to where $u(z)=u_{max}$. For our profiles the obtained shape factors range between $1.9$ to $2.3$, which are smaller than $H=2.59$, the value for a laminar Prandtl-Blasius boundary layer. A shape factor smaller than the theoretical value means the corresponding profile will approach its asymptotic value (the maximum velocity) slower than the theoretical profile does. 

In figure~\ref{fig:H-four-angles} we show the shape factor $H$ for mean velocity profiles obtained in the laboratory and dynamical frames respectively for the four tilting angles and for all $Ra$ measured. The dashed lines in the figure indicate the Prandtl-Blasius value of 2.59. It is seen that, despite the data scatter, there is a general trend that for both lab- and dynamical-frame profiles the deviation from the Prandtl-Blasius profile increases with $Ra$. This is no surprise,  since, as the convective flow above the BL becomes more turbulent with increasing $Ra$, the BL itself will experience stronger fluctuations and hence larger deviations from the laminar case.  This finding that the dynamical rescaling method works better for smaller Ra than larger ones is consistent with those found in DNS studies in the same geometry by \citet{stevens2012pre} for the temperature profile and by \citet{schumacher2012jfm} and by \citet{scheel2012jfm} for the velocity profile. The second feature is that for all $\theta$ and $Ra$ the profiles obtained in the dynamical frame in general show some degree of improvement towards that of Prandtl-Blasius value as compared to those obtained in the laboratory frame. We also note that the ``degree of improvement does not seem to have an obvious dependence on $Ra$, which is also consistent with the findings of \citet{zhou2010prl,zhou2010jfm}. 

Some insight can be obtained by examining the rescaled instantaneous velocity profiles. Figure~\ref{fig:1degree-6E+8-Instantaneous} show examples of rescaled instantaneous velocity profiles, where the distance from the plate has been normalized by the instantaneous BL thickness corresponding to that moment and the velocity has been normalized by the instantaneous maximum horizontal velocity. It is seen that there are quite few cases where the rescaled instantaneous velocity profile is rather close to the theoretical Prandtl-Blasius profile (up to the point of the maximum velocity) and deviations of the instantaneous shape are likely caused by distubances such as plume emissions. Also shown in the figure are the shape factor $H(t)$ of these instantaneous profiles.  To quantify how the instantaneous profiles are distributed with respect to the Prandtl-Blasius profile, we examine the PDF of the shape factor difference $\delta H(t) = H(t)-H^{PB}$ where $H^{PB}=2.59$. Figure~\ref{fig:DeltaH-pdf} plots the PDFs of $\delta H(t)$ for the 4 tilting angles and for all measured $Ra$ respectively. Despite the seemingly large variations among them, these PDFs show the general trend that the rescaled instantaneous profiles measured at lower values of $Ra$ ($\lesssim  1\times 10^{9}$) are more of the time having a shape closer to that of the Prandtl-Blasius profile and that for higher values of $Ra$ the peak of the PDFs shift to smaller values of $H$. This indicates that with increasing $Ra$ the profiles around the BL thickness becomes more rounded, i.e. the approach to the maximum velocity becomes slower and slower.  We further note that these general trends are true across all tilt angles. Another feature observed in the present 3D case is that we did not find any strong correlation between the instantaneous BL thickness $\delta_{v}$ and the velocity $u(t)$ just above the BL. This is in contrast to the finding in the quasi-2D experiment where $\delta_{v}$ and $u(t)$ are found to have a strong negative correlation, i.e. a large velocity above would exert a stronger shear and therefore thins the BL thickness \citep{zhou2010prl}. This result suggest that in certain aspect the BLs in the 3D and in the 2D/quais-2D cases are dynamically different.

\section{Summary and conclusions}\label{sec:summary}
We have conducted an experimental study of velocity boundary layer properties in turbulent thermal convection.  High-resolution two-dimensional velocity field was measured using the particle image velocimetry (PIV) technique in a cylindrical cell of height $H=18.6$ cm and aspect ratio close to unity, with the Rayleigh number $Ra$ varying from $10^{8}$ to $6\times10^{9}$ and the Prandtl number $Pr$ fixed at $\sim 5.4$, with the convection cell tilted with respect to gravity at angles $\theta=0.5^{o}$, $1^{o}$, $2^{o}$, and $3.4^{o}$, respectively. Measurements made with small $\theta$ are aimed at studying BL properties under more steady shear, but the BL itself is assumed to be unperturbed otherwise. For large values of $\theta$ we wish to examine how the BL responds to relatively large perturbations. We also examined effectiveness of the dynamical BL scaling method in a three-dimensional system. 

It is found that the Reynolds number $Re$ ($=U_{max}H/\nu$) based on the maximum mean horizontal velocity scales with $Ra$ as $Re \sim Ra^{0.43}$ and the Reynolds number $Re_{\sigma}$ ($=\sigma_{max}H/\nu$) based on the maximum rms velocity scales with $Ra$ as $Re_{\sigma} \sim Ra^{0.55}$. Both exponents do not seem to have an apparent dependence on the tilt angle. On the other hand, the amplitude of $Re$ seem to show a weak increasing trend with $\theta$. 

With the measured horizontal velocity, we obtain two length scales, i.e. the viscous BL thickness $\delta_{v}$ based on the mean horizontal velocity profile and the length scale $\delta_{\sigma}$ based on the rms horizontal velocity profile. It is found that as far as scaling with the Reynolds number $Re$ is concerned, the behavior of $\delta_{v}$ can be divided into two regimes according to the tilting angle of the cell. For $\theta \le 1^{o}$, it is found that $\delta_{v} \sim Re^{-0.46\pm0.03}$, which within experimental uncertainty may be considered to be consistent with that of the Prandtl-Blasius BL. 
It thus appears that the main effect of tilting the cell is to restrict the azimuthal meandering of the large-scale circulation but the BL is otherwise not strongly perturbed. For $\theta \geq 1^{o}$, the absolute value of the exponent is found to increase with $\theta$ and in this case the BL may be considered to be strongly perturbed.  It is found that the scaling exponent of $\delta_{\sigma}$ with respect to $Ra$ ($Re$) does not have a strong dependence on $\theta$ as $\delta_{v}$ does. But similar to $\delta_{v}$, the absolute values of these exponents increase with increasing $\theta$.
 
It is also found that tilting the cell modifies the velocity profile in the BL region, i.e. for different tilt angles the shape of profiles is different. But for the same tilting angle the velocity profiles measured at different $Ra$ can be brought to collapse on a single curve when the mean velocity is normalized by the maximum velocity $U_{max}$ and the distance from the plate by the viscous BL thickness $\delta_{v}$. 
 
With simultaneously measured horizontal and vertical velocity components, we also obtain the Reynolds stress $\tau_{R}$ in the velocity boundary layer. It is found that $\tau_{R}$ is stronger in the mixing zone comparing with the rectangular cell. The wall quantities such as the wall shear stress$\tau_{w}$, the viscous sublayer$\delta_{w}$, the friction velocity $u_{\tau}$ are also measured. Their scaling exponents with the Reynolds number are very close to those predicted for classical laminar boundary layers, which is also consistent with the measurement in rectangular cell.
 
Regarding the dynamical scaling method, we found that the method in general works better when the cell is tilted at larger angle $\theta$ than it does at smaller angles, but the effect is somewhat marginal. With respect to the influence of $Ra$, it is found that in general the mean velocity profile sampled in both the laboratory and dynamical frames are more closer to the Prandtl-Blasius profile at smaller values of $Ra$ than they are at larger $Ra$, which is consistent with findings from previous DNS  studies. Moreover, it is found that for smaller values of $Ra$ ($\lesssim  1\times 10^{9}$) the PDF's of the shape factor $H$ for the rescaled instantaneous profiles exhibit a peak close to that for the Prandtl-Blasius profile, whereas for larger values of $Ra$ the peaks shift to smaller values of $H$, indicating the profile's approach to the maximum velocity becomes slower and slower with increasing $Ra$. Another finding is that the effectiveness of the dynamical scaling method, in terms of its ability of bringing the mean velocity profile closer to that of Prandtl-Blasius profile, does not have any apparent dependence on $Ra$. Our general conclusion is that as far as the effectiveness of the dynamical scaling method is concerned the influence of titling angle is much smaller  than that of the Rayleigh number $Ra$. We note that the Prandtl-Blasius boundary layer theory is a 2D model, so it is perhaps no surprise that the dynamic method works less well in 3D than in 2D. 
%But the fact that is works to the level it does is XXXX

\appendix

We would like to thank X.-D. Shang and S.-Q. Zhou for kindly making their PIV facility available to us and L. Qu, Y.-C. Xie and S.-D. Huang for helping with the experiment. This work was supported in part by the Hong Kong Research Grants Council under Project Nos. CUHK404409 and CUHK403811.

\clearpage

\clearpage

%\bibliographystyle{jfm}

%\bibliography{JFM-V10}

\end{document}